\documentclass[pra,reprint,onecolumn,aps,notitlepage,nofootinbib,superscriptaddress, preprintnumbers]{revtex4-2}
\usepackage{amsmath}
\usepackage{amssymb}
\usepackage{bm,braket}
\usepackage{qcircuit}
\usepackage{float}
\usepackage{xcolor}
\usepackage{comment}
\usepackage[com pat=1.1.0]{tikz-feynman}
\usepackage{hyperref}
\usepackage{multirow}
\usepackage{subfig}
\def\be{\begin{equation}}
\def\ee{\end{equation}}
\def\bea{\begin{eqnarray}}
\def\eea{\end{eqnarray}}
\def\bes{\begin{eqnarray}}
\def\ees{\end{eqnarray}}
\def\bi{\begin{itemize}}
	\def\ei{\end{itemize}} 
\usepackage{amsthm}
\theoremstyle{definition}

\definecolor{rindou1}{rgb}{0.4431,0.2862,0.7960}
\definecolor{rindou2}{rgb}{0.0078,0.1215,0.4392}
\definecolor{lapis}{rgb}{0.0.0470,0.2941,0.5568}
\definecolor{mn}{rgb}{0.15, 0.35, 0.95}

\hypersetup{%
	colorlinks,
	citecolor=magenta,
	filecolor=BrickRed,
	linkcolor=rindou2,
	urlcolor=blue,
	linktocpage=true
}

\usepackage{tikz}
\usetikzlibrary{braids}

\begin{document}
\title{Quantum computation of SU(2) lattice gauge theory 
with continuous variables}
\author{Victor Ale}
	\email{vale@vols.utk.edu}
\affiliation{Department of Physics and Astronomy, The University of Tennessee, Knoxville, TN 37996-1200, USA}
	\author{Nora M.\ Bauer}
	\email{nbauer1@vols.utk.edu}
\affiliation{Department of Physics and Astronomy, The University of Tennessee, Knoxville, TN 37996-1200, USA}
\author{Raghav G. Jha}
\email{raghav.govind.jha@gmail.com}
\affiliation{Thomas Jefferson National Accelerator Facility, Newport News, VA 23606, USA}
\author{Felix Ringer}
\email{felix.ringer@stonybrook.edu}
\affiliation{Thomas Jefferson National Accelerator Facility, Newport News, VA 23606, USA}
\affiliation{Department of Physics, Old Dominion University, Norfolk, VA 23529, USA}
\affiliation{Department of Physics and Astronomy,
Stony Brook University, New York 11794, USA}
\author{George Siopsis}
	\email{siopsis@tennessee.edu}
\affiliation{Department of Physics and Astronomy, The University of Tennessee, Knoxville, TN 37996-1200, USA}

\preprint{JLAB-THY-24-4217}

\date{\today}
\begin{abstract}
We present a quantum computational framework for SU(2) lattice gauge theory, leveraging continuous variables instead of discrete qubits to represent the infinite-dimensional Hilbert space of the gauge fields. We consider a ladder as well as a two-dimensional grid of plaquettes, detailing the use of gauge fixing to reduce the degrees of freedom and simplify the Hamiltonian. We demonstrate how the system dynamics, ground states, and energy gaps can be computed using the continuous-variable approach to quantum computing. Our results indicate that it is feasible to study non-Abelian gauge theories with continuous variables, providing new avenues for understanding the real-time dynamics of quantum field theories. 
\end{abstract}

\maketitle

\tableofcontents

\onecolumngrid

\newpage 

\section{Introduction}\label{sec:I}

Quantum field theories with local gauge symmetries are central to modern fundamental physics as they form the foundation of the Standard Model of particle physics. The interactions in nature (except gravity) are governed by the $SU(3) \times SU(2) \times U(1)$ gauge symmetry. However, studying gauge theories beyond perturbation theory is challenging due to the lack of analytical tools. In his seminal work~\cite{PhysRevD.10.2445}, Wilson introduced a method to study gauge theories, such as $SU(3)$ gauge theory (the theory of the strong interaction), by discretizing Euclidean spacetime. Since then, these methods commonly referred to as `lattice gauge theories' have enabled a range of precision computations, including the determination of hadron masses. The primary tools for these studies are large-scale Monte Carlo simulations. However, Monte Carlo methods have inherent limitations since they rely on real actions such as those in Euclidean space or non-complex actions. This is due to the nature of the algorithms where the exponential of the Euclidean lattice action is treated as a probability distribution. While certain workarounds exist for specific problems like elastic and low-energy scattering, the computational cost of many interesting problems is expected to be beyond the reach of classical computing. This generally includes real-time dynamics, transport coefficients, out-of-time-order correlators, chaos and thermalization, and finite-density properties of strongly interacting matter that require novel numerical methods. In lower dimensions, it is often possible to use tensor network methods for understanding the real-time dynamics~\cite{Silvi:2014pta, PhysRevX.6.011023} and study complex actions~\cite{Bloch:2021mjw, PhysRevResearch.6.033057} but a general effective tool is missing for higher dimensions, especially in 3+1-dimensions. 

A promising approach to address these challenges lies in large-scale quantum computing using the Hamiltonian formulation of lattice gauge theories. For quantum chromodynamics (QCD), this is a challenging problem due to the field content~\cite{Feynman1981} as emphasized in work by Feynman around the same time when he was thinking about the possibility of quantum computers. The application of quantum computing methods to theories with local gauge symmetries, such as QCD, is the central theme of this paper. However, to simplify the discussion, we limit ourselves to $SU(2)$ gauge theory in one and two spatial dimensions, providing a foundation for future extensions toward QCD. Shortly after Wilson's pioneering work on lattice gauge theory, the Hamiltonian approach to non-Abelian lattice gauge theory was developed by Kogut and Susskind~\cite{Kogut1975}. Since then, the Hamiltonian approach has been studied extensively~\cite{Muller:1983rs, Bronzan1985,Ligterink2000} over the decades exploring different approaches such as the prepotential approach using Schwinger bosons~\cite{Mathur:2004kr, Mathur:2005fb} and its recent extension known as the loop-string-hadron formulation~\cite{Raychowdhury:2019iki}. However, challenges remain due to the hardness of simultaneous non-satisfiability of local dynamics, the correct continuum limit, and discrete quantum numbers. 

With recent advances and the possibility of fault-tolerant quantum computers on the horizon, the Hamiltonian approach to field theories is gaining renewed interest. There has been a substantial amount of work related to the Hamiltonian approach to gauge theories; see Ref.~\cite{Zohar:2012xf,Banerjee:2012xg,Kuhn:2015zqa,Mezzacapo:2015bra,Klco:2019evd,Davoudi:2020yln,ARahman:2021ktn,Ciavarella:2021nmj,Farrell:2022wyt,Davoudi:2022xmb,Alexandru:2023qzd,Muller:2023nnk,DAndrea:2023qnr,Calajo:2024qrc,Mariani:2024osg} and references therein. Several approaches have been developed for digitizing gauge theories, primarily focused on mapping the Hamiltonian to qubits or discrete variables. In this work, we focus instead on the use of continuous variables. The elementary building blocks of continuous-variable quantum computing are qumodes or quantum mechanical harmonic oscillators that have an infinite-dimensional Hilbert space. One of the problems associated with the simulation of gauge theories using qubit-based quantum computers is that the infinite degrees of freedom corresponding to a continuous group symmetry have to be truncated~\cite{Assi:2024pdn}. Different than for fermions, where the degrees of freedom per lattice site are finite, simulating gauge fields with qumodes may therefore be more natural due to their (formally) infinite-dimensional Hilbert space. In this paper, we develop a formulation of the lattice Hamiltonian for $SU(2)$ gauge theory that is well suited for continuous variable quantum computing. We study both the ground and excited state preparation using a coupled cluster ansatz as well as real-time dynamics. We work in the maximal tree gauge, which eliminates a large part of unphysical degrees of freedom. We consider two approaches that require either three or four qumodes per physical link variable. This gauge-fixing procedure was first discussed in Ref.~\cite{Creutz:1976ch} and subsequently followed in several recent studies~\cite{Gattringer:2010zz, Bauer:2023jvw, Carena:2024dzu, Mariani:2024osg, Grabowska:2024emw}. This choice fixes all local gauge symmetries up to a global gauge transformation at the base of the maximal tree.

This work is an extension of our previous work on the continuous variable formulation of the $O(3)$ model \cite{Jha:2023ecu}. In this work, we demonstrated that it is possible to carry out quantum computations of this model using three qumodes per lattice site\footnote{This requirement could be reduced to two qumodes~\cite{Jha:2023ump} due to the requirement that unit vectors take values on a two-sphere.}. Here, we provide the first continuous variable exploration of a local non-Abelian gauge theory. In particular, we focus on open boundary conditions for a one-dimensional ladder or string of plaquettes and a two-dimensional square grid of plaquettes. 

In Ref.~\cite{Bazavov:2019qih}, SU(2) gauge theory in two Euclidean dimensions with matter in unitary gauge was studied using the character expansion around the strong coupling using real-space tensor renormalization group method. This was done in the basis where the electric term of the Hamiltonian is diagonal, also referred to as the irreducible representation of the gauge group basis or simply the ``electric'' basis. However, as we approach the continuum limit, an increasingly large truncation is required. Since the electric and magnetic parts of the Hamiltonian do not commute, there is no common basis where both terms are diagonal. Another option that is often used is the ``magnetic basis'' where the magnetic Hamiltonian is diagonal. In addition, a mixed basis has been explored in Ref.~\cite{Bauer:2023jvw}. 

The paper is structured as follows. In Section \ref{sec:II}, we express the $SU(2)$ gauge theory Hamiltonian in terms of quaternions exploiting the relation that $SU(2) \otimes SU(2) / \mathbb{Z}_{2} \equiv SO(4)$. In Section \ref{sec:III}, we consider the one plaquette case where exact results are known for all values of the coupling. We also introduce the gauge-fixed Hamiltonian in this section for the single plaquette. In Section~\ref{sec:IV}, we study a chain of plaquettes with open boundary conditions and compute the ground state energy density and gap while in Section~\ref{sec:V} we study the extension to two spatial dimensions. In Section~\ref{sec:VI}, we introduce our quantum algorithms based on qumodes for the simulation of $SU(2)$ lattice gauge theory. We end the paper with a brief discussion on future extensions of our work in Section~\ref{sec:VII}.

\section{$SU(2)$ gauge theory on a spatial lattice}\label{sec:II}

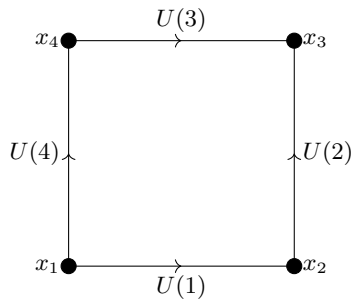
\begin{figure}[t]
\centering
\tikzset{->-/.style={decoration={
  markings,
  mark=at position .5 with {\arrow{>}}},postaction={decorate}}}
\begin{tikzpicture}
    \filldraw (6,2) circle (0.1cm) node[anchor=east]{$x_1$};
    \filldraw (6,5) circle (0.1cm) node[anchor=east]{$x_4$};
    \filldraw (9,2) circle (0.1cm) node[anchor=west]{$x_2$};
    \filldraw (9,5) circle (0.1cm) node[anchor=west]{$x_3$};
    \draw[->-] (6,2) to (6,5); \draw[->-] (6,2) to (9,2); \draw[->-] (9,2) to (9,5); \draw[->-] (6,5) to (9,5);
    \node[anchor = north] at (7.5,2) {$U(1)$};
    \node[anchor = west] at (9,3.5) {$U(2)$};
    \node[anchor = south] at (7.5,5) {$U(3)$};
    \node[anchor = east] at (6,3.5) {$U(4)$};
\end{tikzpicture}
\caption{The single plaquette with links specified. The gauge-invariant Wilson loop expectation value is computed using a path-ordered trace of four links $U$ in the counterclockwise direction. }
    \label{fig:plaquette}
\end{figure}

Gauge theory with a local $SU(2)$ symmetry can be formulated in terms of matrices $U(a)$ residing on a link labeled $a$ along which one also defines the angular momentum operator $J_{\mu\nu} (a)$ ($\mu,\nu = 0,1,2,3$). We parameterize the unitary matrix as $U(a) = u_0(a) \mathbb{I} + i\vec{u} (a)\cdot \vec{\sigma}$, with the Pauli matrices $\vec{\sigma} = (\sigma_1,\sigma_2,\sigma_3)$ and unit quaternions $\bm{u} (a)$ satisfying $\bm{u} (a)\cdot \bm{\bar{u}} (a) \equiv u_0^2 (a) + {\vec{u}}^2(a) = 1$. 
Using the analog of the angular momentum for the rigid rotator, we can write down the kinetic (electric) part of the $SU(2)$ lattice Hamiltonian. The angular momentum at link $a$ is defined as 
\be
\label{eq:ang_mom1}
J_{\mu\nu} (a) = -i \left( u_\mu(a) \frac{\partial}{\partial u_\nu (a)} - u_\nu (a) \frac{\partial}{\partial u_\mu (a)} \right) \,,
\ee
with $\mu,\nu =0,1,2,3$. 
The Hamiltonian discretized on a spatial lattice of arbitrary dimension can be written as
\be 
\bm{H} = g^2 {H} \ , \ \ H =  \frac{1}{2} \sum_{\text{links}\ a} J_{\mu\nu}^2 (a) + \lambda  \sum_{\Box} (1 - W[C]) 
\label{eq:KSHam}
\ee 
where $\lambda = 4/g^{4}$ and the lattice spacing has been set to one. Our conventions follow Ref.~\cite{Kogut1975}. It is convenient to work with the scaled Hamiltonian $H$. Here, $W[C]$ is the Wilson loop over the plaquette $C$ consisting of links labeled $1,2,3,4$ joining lattice sites $x_1,x_2,x_3,x_4$. It is given by 
\be \label{eq:WC}
W[C] = \frac{1}{2} \text{Tr} [U(1) U(2) U^\dagger (3)  U^\dagger (4) ],
\ee 
as shown in Fig.~\ref{fig:plaquette}. The normalization ensures that the ordered trace of the Wilson loop has a maximum value of 1. In the second term Eq.~\ref{eq:KSHam}, we sum over plaquettes. This path-ordered trace of the Wilson loop can be expressed in terms of unit quaternions $\bm{u} (i)$ with $i=1,2,3,4$. In this paper, we consider $D=1, 2$ spatial dimensions. The $D=1$ case is often referred to as a \textit{ladder/string of plaquettes}. The first term in Eq.~\ref{eq:KSHam} involving angular momenta is usually referred to as the electric term and the second term involving plaquettes is called the magnetic term. The continuum limit of this lattice Hamiltonian i.e., the limit where the correlation length $\xi$ is large relative to the lattice spacing ($\xi \sim e^{1/g^2} \gg 1$) is obtained as we take the weak-coupling limit of $\lambda \to \infty$. In this limit, the gauge links $U \to \mathbb{I}$, $u_{0} \to 1$, and $W[C] \to 1 $.   

Since this is a constrained system, the Hamiltonian $H$ does not determine the Hilbert space by itself. The gauge-invariant or physical states $\ket{\Psi}$ obey the following constraint at each link
\be\label{eq:constr} \mathcal{C} (a) \ket{\Psi} = 0 \, ,\; \text{with}\;\; \mathcal{C} (a) \equiv \epsilon^{\mu\nu\rho\sigma} J_{\mu\nu} (a) J_{\rho\sigma} (a) \,. \ee 
The electric part of $H$ and all $\mathcal{C} (a)$ are Casimir operators commuting with all components of the angular momentum $J_{\mu\nu} (a)$. The Wilson loop
$W[C]$ also commutes with the constraint in Eq.~\eqref{eq:constr}: $[\mathcal{C} (a) , W[C]] = 0$ ($a=1,2,3,4$).
Thus, for a given state $\ket{\Psi}$ that satisfies the constraint in Eq.~\eqref{eq:constr}, e.g. the state $\ket{\Psi}=W[C] \ket{0}$, we obtain a new state by acting on it with the angular momentum operator $J_{\mu\nu} (a)\ket{\Psi}$ that also satisfies the constraint Eq.~\eqref{eq:constr}.

Additionally, the system obeys Gauss's law, which further constrains the Hilbert space to the gauge singlet sector \cite{Kogut1975}. To define it, note that the four-dimensional angular momentum algebra splits into two three-dimensional angular momentum algebras, $\vec{L}^{R,L}$, that commute with each other\footnote{This is referred to as body-fixed and space-fixed reference frames of the rigid rotator in Ref.~\cite{Kogut1975}.}. We have
\be \label{eq:ang_mom} L_1^{R,L} (a) = \frac{1}{2} (J_{01}(a) \pm J_{23}(a) ) \ , \ \ L_2^{R,L}(a) = \frac{1}{2} (J_{02}(a) \pm J_{31}(a) ) \ , \ \ L_3^{R,L}(a) = \frac{1}{2} (J_{03}(a) \pm J_{12}(a) ) \,.
\ee
Gauss's law at lattice site $x$ reads
\be\label{eq:Gauss} \mathbb{G} (x) \ket{\Psi} = 0 \,,\;\; \text{with}\;\;  \mathbb{G} (x) \equiv \sum_{a\ \mathrm{\textcolor{blue}{starts\ at}\ x}} \vec{L}^L (a) + \sum_{a\ \mathrm{\textcolor{blue}{ends\ at}\ x}} \vec{L}^R (a) \,. \ee

\section{Single plaquette}
\label{sec:III}
We start with the one plaquette system where an exact analytic solution can be obtained \cite{PhysRevD.31.3201}. The plaquette consists of links $a=1,2,3,4$ joining lattice sites $x=x_1,x_2,x_3,x_4$, as shown in Fig.~\ref{fig:plaquette}. Gauss's law Eq.~\eqref{eq:Gauss} amounts to the constraints
\be\label{eq:68} \big(\vec{L}^{R}(1) + \vec{L}^{L} (2) \big) \ket{\Psi} = \big( \vec{L}^{R} (2) + \vec{L}^{L} (3) \big) \ket{\Psi} = \big( \vec{L}^{R} (3) + \vec{L}^{L} (4) \big) \ket{\Psi} = \big( \vec{L}^{R} (4) + \vec{L}^{L} (1) \big) \ket{\Psi} = 0 \,. \ee
To construct the Hilbert space, note that on a given link, $\vec{L}^{R}$ and $\vec{L}^{L}$ commute with each other, thus forming independent algebras due to 
$\mathfrak{so}(4) \equiv \mathfrak{su}(2) \oplus \mathfrak{su}(2)$. We can write their common eigenstates as $\ket{jm_Lm_R} \equiv \ket{jm_L}_L \ket{jm_R}_R$ ($m_L,m_R = -j, \dots , j$), where we suppressed the link index. For the pure gauge theory we are considering here, Gauss's law restricts the Hilbert space to the vacuum charge sector, leading to the following basis of states
\be \bigotimes_{a=1}^4 \ket{j_{a}m_{aL}m_{aR}}_{a} \, , \, \ket{j_{a}m_{aL}m_{aR}}_{a} = \ket{j_am_{aL}}_{aL} \ket{j_am_{aR}}_{aR}  \,. \ee
As an example, consider the case of spin-1/2 states $j_a=1/2$ with $m_{aL,R}=\pm 1/2$, for $a=1,2,3,4$.
To simplify the notation, we will omit the quantum numbers $j_a$ and write $m_{aL,R} = \pm$. Using this notation, the basis states of the Hilbert space are $\bigotimes_{a=1}^4 \ket{m_{aL}}_{aL} \ket{m_{aR}}_{aR}$.
First, let us focus on the states $\ket{m_{1R}}_{1R} \ket{m_{2L}}_{2L}$. The total spin needs to be zero due to Gauss's law Eq.~\eqref{eq:68}. Therefore, we need to pick the singlet state, which is given by the linear combination
\be \label{eq:75} \frac{1}{\sqrt{2}} \left( \ket{+}_{1R} \ket{-}_{2L} - \ket{-}_{1R} \ket{+}_{2L} \right)\,,
\ee
satisfies the first constraint of Gauss's law. We can choose the same singlet states for the other pairs of states to satisfy the remaining three constraints. Thus  we arrive at the state
\be \frac{1}{4} \left( \ket{+}_{1R} \ket{-}_{2L} - \ket{-}_{1R} \ket{+}_{2L} \right) \left( \ket{+}_{2R} \ket{-}_{3L} - \ket{-}_{2R} \ket{+}_{3L} \right) \left( \ket{+}_{3R} \ket{-}_{4L} - \ket{-}_{3R} \ket{+}_{4L} \right) \left( \ket{+}_{4R} \ket{-}_{1L} - \ket{-}_{4R} \ket{+}_{1L} \right) \,,
\ee
which satisfies Gauss's law Eq.~\eqref{eq:68} by construction. This state can be verified to be  $2W[C]\ket{\Omega}$, where $\ket{\Omega}$ is the ground state of the electric part of the Hamiltonian with zero angular momentum ($j=0$, see Eq.~\eqref{eq:14H} below). The overall factor can be deduced from the normalization constraint $\| 2W[C] \ket{\Omega} \|^2 = 4 \bra{\Omega} [W[C]]^2 \ket{\Omega} = 1$, using  Eq.~\eqref{eq:WC}. Similarly, it is verified that using the singlet state in the $j=1$ subspace 
\be \frac{1}{\sqrt{3}} (\ket{-1}_{1R}\ket{1}_{2L} - \ket{0}_{1R}\ket{0}_{2L} + \ket{1}_{1R}\ket{-1}_{2L}) \,,
\ee 
can be written as $(4[W[C]]^2 - 1)\ket{\Omega}$, which is orthogonal to $\ket{\Omega}$. The overall constant is deduced again from the normalization $\| (4[W[C]]^2 - 1)\ket{\Omega} \|^2 = 1$. In the $j = \frac{3}{2}$ subspace, the singlet state is given by 
\be \frac{1}{2} \Big(\ket{-\frac{3}{2}}_{1R}\ket{\frac{3}{2}}_{2L} - \ket{-\frac{1}{2}}_{1R}\ket{\frac{1}{2}}_{2L} + \ket{\frac{1}{2}}_{1R}\ket{-\frac{1}{2}}_{2L} - \ket{\frac{3}{2}}_{1R}\ket{-\frac{3}{2}}_{2L}\Big) \,,
\ee which matches $(8[W[C]]^3 - 4W[C]) \ket{0}$. It is orthogonal to $2W[C]\ket{0}$ and the overall constant is such that $\| (8[W[C]]^3 - 4W[C])\ket{\Omega} \|^2 = 1$. Thus the Hilbert space can be constructed in terms of singlet vectors in the various $j$-subspaces \be\label{eq:14H} \mathcal{H} = \{\ket{0}_{j=0}, \ket{0}_{j=\frac{1}{2}}, \ket{0}_{j=1}, \ket{0}_{j=\frac{3}{2}}, \dots \ket{0}_{j=j_{\text{max}}/2}\} = \{ \ket{\Omega}, 2W[C]\ket{\Omega}, (4[W[C]]^2 - 1)\ket{\Omega}, (8[W[C]]^3 - 4W[C])\ket{\Omega}, \dots \}\,.\ee
In this basis, the electric part of the Hamiltonian is diagonal, and it is a convenient choice in the small $\lambda$ limit in which the electric part dominates over the magnetic part. However, it does not capture the correct continuum limit that one obtains for large $\lambda$. We therefore focus on a different approach that does not rely on the electric basis Eq.~\eqref{eq:14H}.

It is known that the single-plaquette system consists of a single variable, which can be defined to be the Wilson loop of the single-plaquette $W[C]$. For the generalization to systems with more than one plaquette, it is convenient to fix the gauge. We can use Gauss's law to fix three of the links to be the identity.
This procedure will be discussed in more detail later and is known as ``maximal tree'' gauge fixing~\cite{Creutz:1976ch}. We introduce the $SU(2)$ matrix 
\be X = U(1) U(2) U^\dagger (3) U^\dagger (4) = x_0 \mathbb{I} + i \vec{x}\cdot \vec{\sigma}\,,\ee
where $x_0^2 + \vec{x}^{\, 2} = 1$. Then $W[C] = \frac{1}{2} \mathrm{Tr} X = x_0$, and the Hamiltonian in Eq.~\eqref{eq:KSHam} reduces to
\be\label{eq:H1pl} H = \frac{1}{2} ( \vec{L}^2 + \vec{K}^2 ) + \lambda (1- x_0) \,,\ee
where $\vec{L}, \vec{K}$ act on $\bm{x} = (x_0, \vec{x})$ with angular momentum and boost operators $L_i = \frac{1}{2} \epsilon^{ijk} J_{jk}$, $K_i = J_{0i}$. To fix the gauge, we only need 3 of the 4 equations in Eq.~\eqref{eq:Gauss}. Imposing the fourth one amounts to further restricting the Hilbert space to states with zero angular momentum $\vec{L} = \vec{0}$, i.e. states with spherically symmetric wave functions $\Psi (x_0)$. Changing coordinates to $x_0 = \cos\chi$, the
energy eigenfunctions obey the Schr\"odinger equation \cite{PhysRevD.31.3201}
 \be\label{eq:A1} - \frac{1}{2\sin^2 \chi} \frac{d}{d\chi} \left[ \sin^2\chi \frac{d\Psi_n (\chi)}{d\chi} \right] + \lambda (1 - \cos\chi) \Psi_n (\chi) = E_n \Psi_n (\chi) \,.\ee
 Here, $\Psi_n$ is periodic with $\Psi_n (\chi + 2\pi) = \Psi_n (\chi)$. Exact solutions are given in terms of \emph{Mathieu functions} and the energy levels $E_n$ are obtained by imposing the periodicity of the solutions. In the weak coupling (large $\lambda$) limit, obtained by letting $x_0 \to 1$, the Hamiltonian simplifies. Approximating $x_0 \approx 1 - \frac{1}{2} \vec{x}^{\, 2}$, the Hamiltonian can be written as
\be \label{eq:15}
H = \sqrt{\lambda} H^{(0)} + \mathcal{O}(\lambda^0) \ , \ \ H^{(0)} = \frac{1}{2} \vec{p}^{\, 2} + \frac{1}{2} \vec{x}^{\, 2} \ , \ \ [ x_i , p_j ] = i\delta_{ij} \ ,
\ee
which describes a three-dimensional harmonic oscillator of frequency $\omega = \sqrt{\lambda}$. We have scaled $\vec{x}$ so that $H^{(0)}$ is independent of $\lambda$. For large $\lambda$, since $x_0 \to 1$ and $\sin \chi \approx \chi$, we can expand Eq.~\eqref{eq:A1} in $\chi$ and obtain to leading order
\be - \frac{1}{2\chi^2} \frac{d}{d\chi} \left[ \chi^2 \frac{d\Psi_n^{(0)} (\chi)}{d\chi} \right] + \frac{1}{2} \chi^2 \Psi_n^{(0)} (\chi) = E_n^{(0)} \Psi_n^{(0)} (\chi)\,. \ee
The solutions involve Hermite polynomials and can be written as
\be E_n^{(0)} = 2n + \frac{3}{2}  \ , \ \ \Psi_n^{(0)} (\chi) \propto \frac{1}{\chi} H_{2n+1} (  \chi) e^{- \frac{1}{2} \chi^2}. \ee
Notice that this list does not include all energy levels of a three-dimensional harmonic oscillator, because only spherically symmetric states are allowed by Gauss's law (with angular momentum quantum number $l=0$).
The wave functions of the ground and first-excited states are given by
\be\label{eq:35} \Psi_0^{(0)} \propto e^{- \frac{1}{2} \chi^2} \ , \ \ \Psi_1^{(0)} \propto (\chi^2 - \frac{3}{2}) e^{- \frac{1}{2} \chi^2} \,, \ee 
with energies $E_0 = \frac{3}{2} \sqrt{\lambda} + \mathcal{O} (\lambda^0)$, $E_1 = \frac{7}{2} \sqrt{\lambda} + \mathcal{O} (\lambda^0)$, respectively. The energy gap is given by 
\be\label{eq:22}\Delta E = E_1 - E_0 = 2\sqrt{\lambda} + \mathcal{O}  (\lambda^0) \ . \ee 
This Hamiltonian can also be obtained without gauge fixing. It is instructive to discuss the details to generalize it to cases involving a large number of plaquettes where gauge fixing can be more cumbersome. Using $\sum J_{\mu\nu}^2 (a) \approx \frac{1}{4}\vec{P}^2 (a)$, where $\vec{P} (a)$ is the momentum conjugate to $\vec{u} (a)$ defined on link $a$, the electric part of the Hamiltonian Eq.~\eqref{eq:KSHam} for a single plaquette in the large $\lambda$ limit is given by
\be\label{eq:17} H_E = \frac{1}{8}  \sum_{a=1}^4 [\vec{P}(a)]^2 \,,\ee
where $[u_i(a), P_j (b)] = i\delta_{ab} \delta_{ij}$. The
magnetic part is given by
\be H_B = 2\lambda (1 - W[C]) \approx \lambda \left[ {\begin{array}{cccc}
   \vec{u} (1), &
   \vec{u} (2), &
   \vec{u} (3), &
    \vec{u} (4) 
  \end{array} } \right] V_B   \left[ {\begin{array}{c}
   \vec{u} (1) \\
   \vec{u} (2) \\
   \vec{u} (3) \\
    \vec{u} (4) 
  \end{array} } \right] \ , \ \ V_B = \frac{1}{4} \left[ {\begin{array}{cccc}
    1 & 1 & -1 & -1 \\
    1 & 1 & -1 & -1  \\
    -1 & -1 & 1 & 1  \\
    -1 & -1 & 1 & 1 
  \end{array} } \right]\,.
\ee
The matrix $V_B$ only has one non-vanishing eigenvalue, $1$, which corresponds to the normalized eigenvector $\bm{e} = \frac{1}{2}[ 1,\ 1,\ -1, \ -1]^T$. This leads to the normal mode
\be \vec{x} = \frac{1}{2} \left( \vec{u} (1) + \vec{u} (2) - \vec{u} (3) - \vec{u} (4) \right)\,, \ee
which matches the result above obtained after gauge fixing.
Gauge-invariant wave functions only depend on $\vec{x}$, and the Hamiltonian reduces to the expression in Eq.~\eqref{eq:15}.

\begin{figure}
    \centering
    \includegraphics[width=0.49\linewidth]{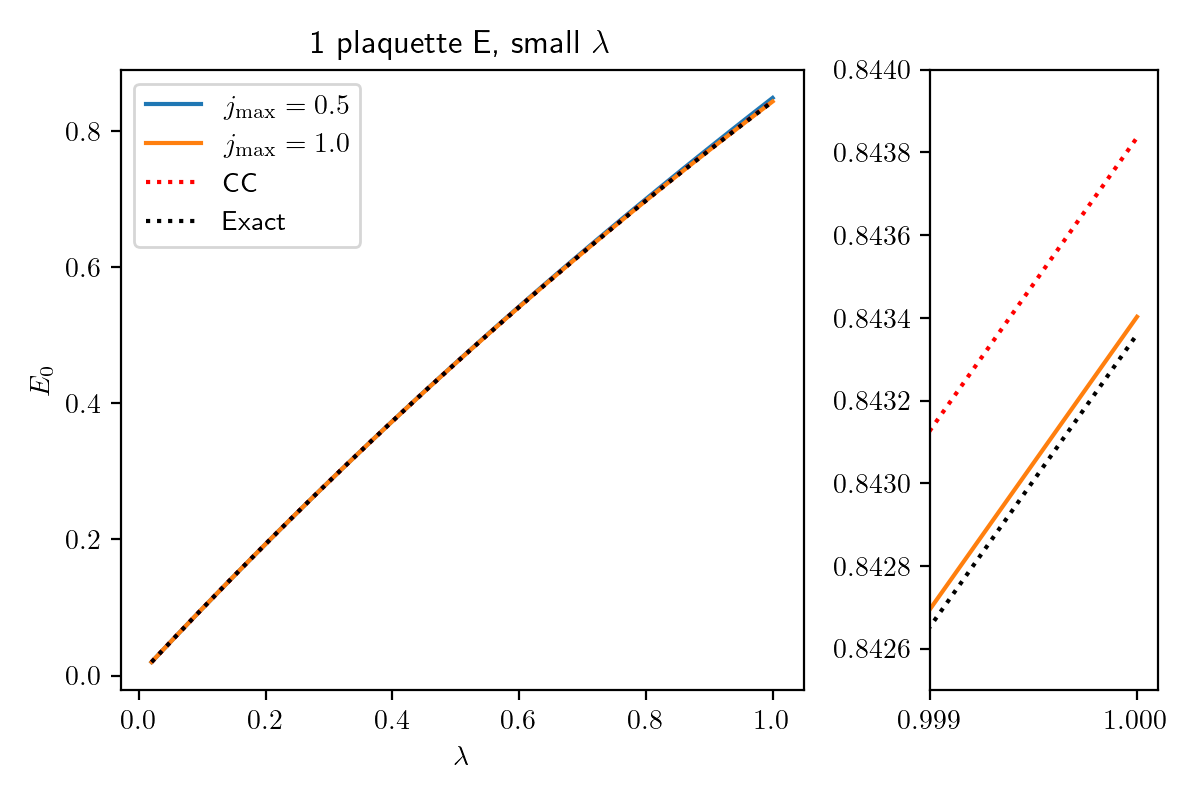}
    \includegraphics[width=0.49\linewidth]{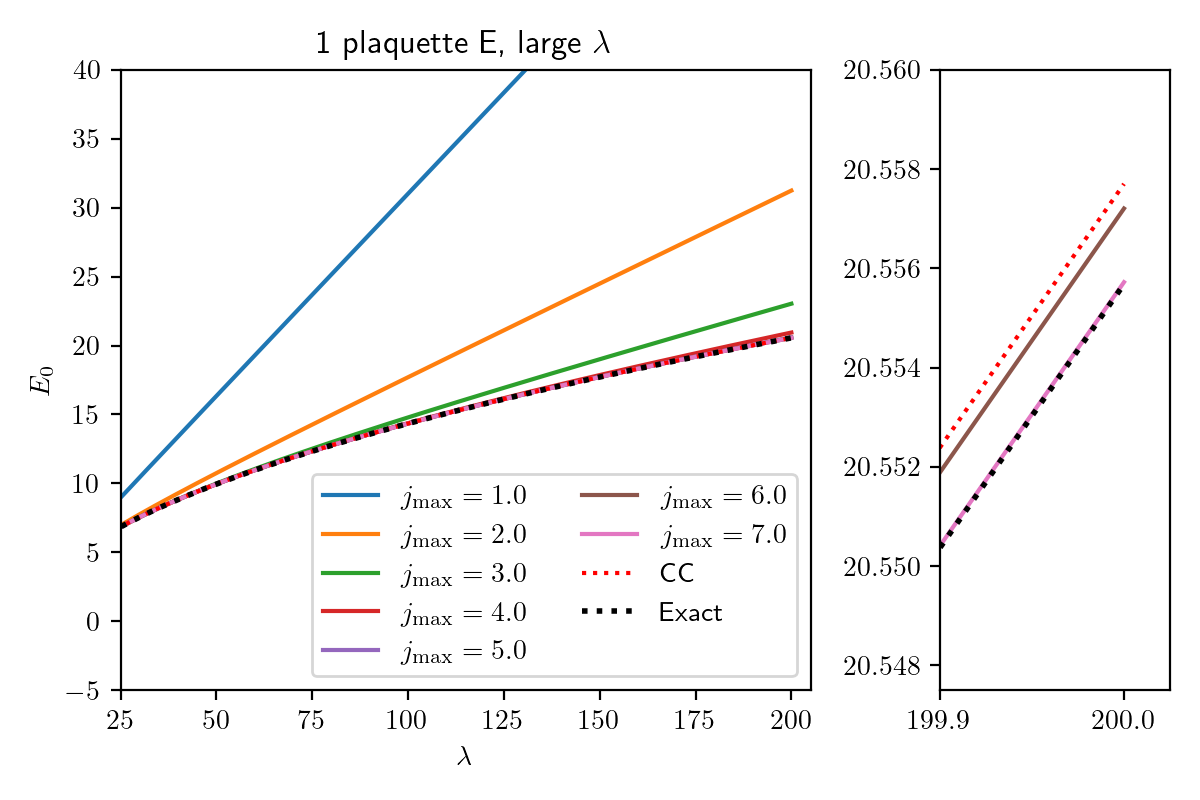}
    \caption{The ground state energy using the coupled cluster ansatz compared with the results using an exact diagonalization of the Hamiltonian in the electric basis with different truncations $j_{\text{max}}$. Note that the eigenvalues of the electric part of the Hamiltonian is $j(j+1)$ for each link where $j = 0, 1/2, 1, \cdots, j_{\text{max}}$. In practice, we fix each link to some maximum allowed value of $j_{\text{max}}$.}
    \label{fig:fig10d}
\end{figure}
\begin{figure}
    \centering
    \includegraphics[width=0.49\linewidth]{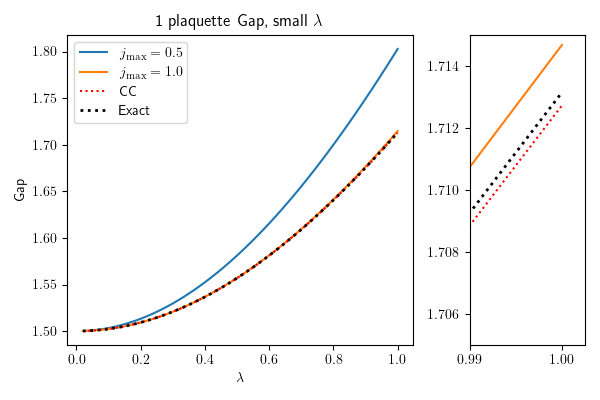}
    \includegraphics[width=0.49\linewidth]{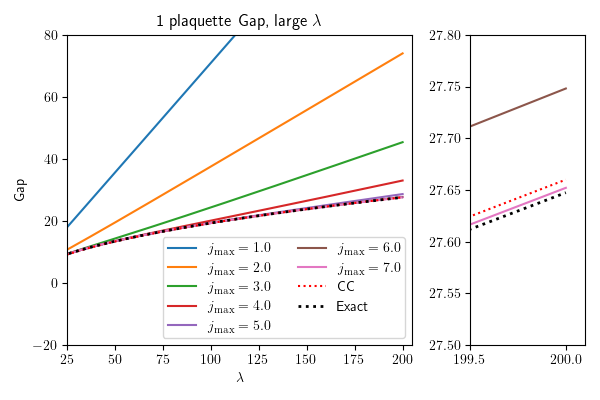}
    \caption{Energy gap for one plaquette using the coupled cluster ansatz compared to exact diagonalization results. We show the results separately for small (left panels) and large $\lambda$ (right panels).}
    \label{fig:fig10e}
\end{figure}

For arbitrary coupling strength, away from the weak coupling approximation considered above, the energy gap can be obtained with a variational method. For the ground state, consider the coupled cluster ansatz
\be\label{eq:23} \braket{\bm{x} | \psi_0 (\alpha)} \propto e^{\alpha W[C]} \ , \ee
where $\alpha$ is a variational parameter. The corresponding energy is given in terms of modified Bessel functions of first kind 
\be\label{eq:24} \epsilon_0(\alpha) = \bra{\psi_0 (\alpha)} H \ket{\psi_0 (\alpha)} = \lambda+   \frac{(3 \alpha - 4 \lambda) I_2( 2 \alpha)}{4I_1(2 \alpha)}\,, \ee
where we used the exact expression Eq.~\eqref{eq:H1pl} for the Hamiltonian.
For large $\lambda$, we have $\epsilon_0 (\alpha) \approx \frac{3}{4} (\alpha + \frac{\lambda}{\alpha })$, which is minimized for $\alpha = \sqrt{\lambda}$. We obtain the estimate of the ground state energy $\epsilon_0 (\sqrt{\lambda}) \approx \frac{3}{2} \sqrt{\lambda}$, in agreement with our earlier result.
Similarly, for the first excited state, consider the ansatz
\be\label{eq:25a} \ket{\psi_1 (\alpha)} \propto ( W[C] - \beta) \ket{\psi_0 (\alpha)}\,. \ee 
The parameter $\beta$ is fixed by demanding orthogonality to the ground state. We obtain
\be \beta = \frac{I_2(2\alpha )}{I_1(2\alpha)} \,. \ee
After computing the energy of this trial state, $\epsilon_1(\alpha) = \bra{\psi_1 (\alpha)} H \ket{\psi_1 (\alpha)}$, we obtain the energy gap 
\begin{multline}\label{eq:25e} \Delta E = \epsilon_1 (\alpha) - \epsilon_0(\alpha) \\ = \frac{-2 \left(\alpha^3 + 2\alpha ^2-6\right) \beta +2 (3\alpha + 2 )\alpha ^2 \beta^3 -6 \alpha  +3 (3\alpha +4) \alpha  \beta^2}{4 \alpha   \left(2 \alpha  -3 \beta -2 \alpha  \beta^2 \right)}   + \lambda \frac{ 2 \left(2\alpha ^2-3\right) \beta -4 \alpha ^2 \beta^3 +3 \alpha  -9
   \alpha  \beta^2 }{\alpha   \left(2 \alpha  -3 \beta -2
   \alpha  \beta^2\right)} \end{multline}
where $\alpha$ is chosen to minimize $\epsilon_0(\alpha)$, i.e. solving $d\epsilon_0/d\alpha = 0$.
   
For large $\lambda$, $\epsilon_0$ is minimized for $\alpha = \sqrt{\lambda}$. We obtain $\beta = 1-\frac{3}{4\sqrt{\lambda}} + \mathcal{O} (1/\lambda)$, and we recover the result Eq.~\eqref{eq:22} for the energy gap. For general coupling, we obtained the corresponding results numerically. In Figures \ref{fig:fig10d} and \ref{fig:fig10e}, we compare our results using the coupled cluster (CC) ansatz with exact results, which can be obtained analytically for a single plaquette, as well as results from exact diagonalization for the ground state energy and the energy gap, respectively. Following Refs.~\cite{Klco:2019evd,Rahman2021}, we obtained results from exact diagonalization using different levels of truncation in the electric basis; see original papers for details. The results obtained using the coupled cluster ansatz are in good agreement with exact results for all values of $\lambda$. Agreement with exact diagonalization is also good, but, as expected a higher cutoff is needed at large $\lambda$: For example, $j_{\mathrm{max}} = 1$ suffices at $\lambda = 1$, but we need $j_{\mathrm{max}} = 7$ at $\lambda = 200$ to converge to the exact result as shown in the right inset of Fig.~\ref{fig:fig10d}. 

\section{Ladder of plaquettes}
\label{sec:IV}

Next, we consider a ladder of $N$ plaquettes where we impose open boundary conditions. The results can be easily extended to periodic boundary conditions. We will not discuss periodic boundary conditions in detail here as they do not introduce any significant changes to the relevant physics. 

We start with a system of two plaquettes with open boundary conditions as illustrated in Figure~\ref{fig:twoplaquettes}. In this case, we can use Gauss's law to fix five of the seven gauge links. Thus, the system is described by two $SU(2)$ matrices, which can be conveniently chosen as~\cite{Bauer:2023jvw}
\be X(1) = U(4)U(3)U(2)^\dagger U(1)^\dagger \ , \ \ X(2) = U(1) U(5) U(7) U(6)^\dagger U(2)^\dagger U(1)^\dagger \,,\ee
along the closed paths $C_1\!: x_1\to x_4 \to x_3 \to x_2 \to x_1$ and $C_2\!: x_1\to x_2 \to x_5 \to x_6 \to x_3\to x_2 \to x_1$, respectively. The corresponding Wilson loops are
\be W[C_i] = \frac{1}{2} \mathrm{Tr} X(i) = x_0 (i) \ , \ \ i=1,2 \,. \ee
Here, we used five out of the six equations in Gauss's law (one for each lattice site). The remaining equation imposes the constraint of zero total angular momentum on the states. The Hamiltonian becomes
\begin{equation}
\label{eq:2pq1a}
    H =  \frac{1}{2}\Big( [\vec{L} (1)]^2 +  [ \vec{K} (1) ]^2 + 2[\vec{L} (2) ]^2 + [ \vec{K} (2) ]^2    + \frac{1}{2} (\vec{K} (2) + 3\vec{L} (2)) \cdot (\vec{K} (1) - \vec{L} (1)) \Big) +\lambda(2-x_0(1) - x_0(2))\,,
   \end{equation}
where $L_i (a) = \frac{1}{2} \epsilon^{ijk} J_{jk} (a)$, $K_i(a) = J_{0i}(a)$ analogous to Eq.~\eqref{eq:H1pl} above. In the large $\lambda$ limit, the Hamiltonian reduces to
\begin{equation}
\label{eq:2pq1b}
    H = \sqrt{\lambda} H^{(0)} + \mathcal{O} (\lambda^0) \ , \ \ H^{(0)} =   \frac{1}{2}\Big( [\vec{p} (1)]^2  + [\vec{p} (2) ]^2     + \frac{1}{2} \vec{p} (1) \cdot \vec{p} (2)  \Big) +\frac{1}{2} \left( [\vec{x}(1)]^2 + [\vec{x} (2)]^2 \right) \,,
   \end{equation}
with $[ x_i (a) ,p_j (b)] = \delta_{ab} \delta_{ij}$. This Hamiltonian describes two coupled three-dimensional harmonic oscillators. 
It can be diagonalized by introducing the normal modes $\vec{x}_\pm = \frac{1}{\sqrt{2}} \left[ \vec{x} (1) \pm \vec{x} (2) \right]$ with normal frequencies  $\omega_- = \frac{\sqrt{3\lambda}}{2}$ and $\omega_+ = \frac{\sqrt{5\lambda}}{2}$. 

\tikzset{->-/.style={decoration={
  markings,
  mark=at position .5 with {\arrow{>}}},postaction={decorate}}}
\begin{figure}
\begin{center}
\begin{tikzpicture}
    \filldraw (6,2) circle (0.1cm) node[anchor=east]{$x_1$};
    \filldraw (6,5) circle (0.1cm) node[anchor=east]{$x_4$};
    \filldraw (9,2) circle (0.1cm) node[anchor=north]{$x_2$};
    \filldraw (9,5) circle (0.1cm) node[anchor=south]{$x_3$}; 
    \filldraw (12,2) circle (0.1cm) node[anchor=west]{$x_5$}; 
    \filldraw (12,5) circle (0.1cm) node[anchor=west]{$x_6$}; 
    \draw[->-] (6,2) to (6,5); \draw[->-] (6,2) to (9,2); \draw[->-] (9,2) to (9,5); \draw[->-] (6,5) to (9,5);\draw[->-] (9,5) to (12,5);  \draw[->-] (9,2) to (12,2);    \draw[->-] (12,2) to (12,5); 
    \node[anchor = north] at (7.5,2) {$U(1)$};
    \node[anchor = west] at (9,3.5) {$U(2)$};
    \node[anchor = south] at (7.5,5) {$U(3)$};
    \node[anchor = east] at (6,3.5) {$U(4)$}; 
    \node[anchor = north] at (10.5,2) {$U(5)$}; 
    \node[anchor = west] at (12,3.5) {$U(7)$}; 
    \node[anchor = south] at (10.5,5) {$U(6)$}; 
\end{tikzpicture} 
\caption{The two plaquette ($N=2$) string model with open boundary conditions. An arbitrary $N$ plaquette string has $3N+1$ links. We can set $2N+1$ links to arbitrary $SU(2)$ group element (usually the identity matrix) and hence describe the dynamics using only $N$ physical links as in~\eqref{eq:wilson_loop_2plaq}. 
\label{fig:twoplaquettes}}
\end{center}
\end{figure}
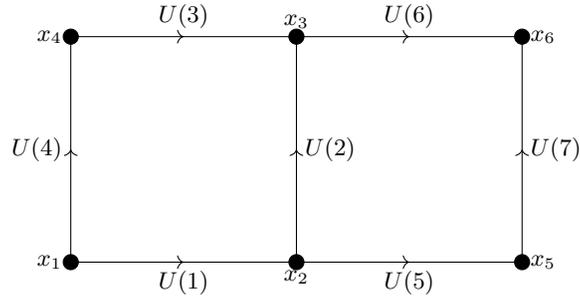

Interestingly, the expression Eq.~\eqref{eq:2pq1b} can be derived in a straightforward manner without gauge fixing, by extending the method discussed in the case of a single plaquette. Note that in the large-$\lambda$ limit, the electric part of the Hamiltonian is diagonal (given by Eq.~\eqref{eq:17}, but with $a=1,\dots,7$), and the magnetic part is given by
\be H_B = \lambda (1 - W[C]) \approx \lambda \vec{\bm{u}}^T \cdot  V_B  \vec{\bm{u}} \ , \ \ V_B = \frac{1}{4} \left[ {\begin{array}{ccccccc}
    1 & 1 & -1 & -1 & 0 & 0 & 0 \\
    1 & 2 & -1 & -1 & -1 & 1 & -1 \\
    -1 & -1 & 1 & 1 & 0 & 0 & 0  \\
    -1 & -1 & 1 & 1  & 0 & 0 & 0 \\
    0 & -1 & 0 & 0 & 1 & -1 & 1 \\
    0 & -1 & 0 & 0 & 1 & -1 & 1 \\
    0 & -1 & 0 & 0 & 1 & -1 & 1 
  \end{array} } \right]  \ , \ \vec{\bm{u}} = \left[ {\begin{array}{c}
   \vec{u} (1) \\
   \vec{u} (2) \\
   \vec{u} (3) \\
    \vec{u} (4) \\
   \vec{u} (5) \\
   \vec{u} (6) \\
    \vec{u} (7)
  \end{array} } \right] \,.\ee
The non-zero eigenvalues of $V_B$ are $\lambda_- = \frac{5}{4}$, $\lambda_+ = \frac{3}{4}$ with corresponding normal mode coordinates
\bea \vec{v}_- &=& \frac{1}{\sqrt{10}} \left[ \vec{u} (1) +2 \vec{u} (2) - \vec{u} (3) - \vec{u} (4) - \vec{u} (5) + \vec{u} (6) - \vec{u} (7) \right] \,,\nonumber\\
 \vec{v}_+ &=& \frac{1}{\sqrt{6}} \left[ \vec{u} (1) - \vec{u} (3) - \vec{u} (4) + \vec{u} (5) - \vec{u} (6) + \vec{u} (7) \right] \,. \eea
Gauge-invariant wave functions only depend on $\vec{x}_\pm$, and we recover the expression Eq.~\eqref{eq:2pq1b} for the Hamiltonian. 

The wave functions for the ground and first excited states, respectively, are given by
\be \Psi_0^{(0)}  \propto e^{- \frac{1}{2} {\omega_+}\vec{v}_+^2 + {\omega_-} \vec{v}_-^2} \ , \qquad \Psi_1^{(0)}  \propto (\omega_- \vec{v}_-^{\, 2} - 3) \Psi_0^{(0)} \,,\ee
with energies 
\be E_0 = \frac{3}{2} (\omega_+ + \omega_-) = 2.976\sqrt{\lambda} + \mathcal{O} (\lambda^0) \ , \ \ E_1 = \frac{3}{2} \omega_+ + \frac{7}{2} \omega_- \approx 4.708\sqrt{\lambda}  + \mathcal{O} (\lambda^0) \,. \ee
Therefore, the gap is 
\be 
\Delta E = 2\omega_- = \sqrt{3\lambda}  + \mathcal{O} (\lambda^0) \,. 
\label{eq:2plaq_gap}
\ee
Away from the weak coupling limit, the energy gap can be obtained with a variational method. For the ground state, consider the coupled cluster ansatz
\be\label{eq:G2} \braket{\bm{x} (1) , \bm{x} (2) | \psi_0 (\alpha)} \propto e^{\alpha ( W[C_1] + W[C_2])} \,. \ee
The corresponding energy that depends on the variational parameter $\alpha$ is given by
\be\label{eq:37a} \epsilon_0 (\alpha) = \bra{\psi_0 (\alpha)} H \ket{\psi_0 (\alpha)} = \frac{3 \alpha  I_2(2 \alpha )}{2 I_1(2 \alpha )} + 2\lambda \left(1 - \frac{I_2(2 \alpha )}{I_1(2 \alpha )} \right) \,. \ee
Similarly, for the excited state, consider the ansatz
\be\label{eq:G21} \ket{\psi_1 (\alpha)} \propto (  W[C_\partial] - \beta ) \ket{\psi_0 (\alpha)}\,, \ee 
where $C_\partial$ traces the perimeter of the lattice. The Wilson loop may be written as
\be 
W[C_\partial] = \frac{1}{2} \mathrm{Tr} [ X(1) X^{\dagger}(2)] = \bm{x}(1)\cdot \bm{x}(2) \,.
\label{eq:wilson_loop_2plaq}
\ee
The parameter $\beta$ is fixed by demanding orthogonality to the ground state Eq.~\eqref{eq:G2}. We obtain
\be \beta = \frac{\bra{\Omega} W[C_\partial] e^{2\alpha ( W[C_1] + W[C_2])} \ket{\Omega}}{\bra{\Omega}  e^{2\alpha ( W[C_1] + W[C_2])} \ket{\Omega}} = \left( \frac{I_2(2\alpha ) }{I_1(2\alpha )} \right)^2 \,.\ee
The energy of the excited state $\epsilon_1 (\alpha) = \bra{\psi_1 (\alpha)} H \ket{\psi_1 (\alpha)}$ can also be obtained exactly in terms of modified Bessel functions.

At large $\lambda$, the coupled cluster ansatze in Eq.~\eqref{eq:G2} and Eq.~\eqref{eq:G21} read, respectively,
\be \braket{\vec{x} (1), \vec{x} (2) | \psi_0^{(0)} (\alpha)} \propto e^{- \frac{\alpha}{2} \left( [\vec{x}(1)]^2 + [\vec{x} (2)]^2 \right) } \ , \ \ket{\psi_1^{(0)} (\alpha)} \propto \left( \frac{1}{2} (\vec{x}_1 - \vec{x}_2)^2 -\frac{3}{2\alpha }  \right) \ket{\psi_0 (\alpha)} \,,\ee
where we used $\beta \approx 1 - \frac{3}{2\alpha }$. Their energies are found to be
\be \label{eq:49a} \epsilon_0^{(0)} (\alpha) = \frac{3}{2} \left( \alpha + \frac{1}{\alpha} \right) \ , \ \ \epsilon_1^{(0)} (\alpha) = \frac{45 \alpha ^{4}+53 \alpha ^{2}  +4 \alpha +50 }{20 \alpha ^{3}-8\alpha ^{2} +20 \alpha   }\ee
The ground-state energy is minimized for $\alpha = 1$. We obtain  $E_0 =\epsilon_0^{(0)}\sqrt{\lambda} + \mathcal{O} (\lambda^0) =  3\sqrt{\lambda} + \mathcal{O} (\lambda^0)$. The excited state energy becomes $E_1 = \epsilon_1^{(0)} (1) \sqrt{\lambda} + \mathcal{O} (\lambda^0) =   4.75\sqrt{\lambda} + \mathcal{O} (\lambda^0)$, and the gap is $\Delta E = 1.75\sqrt{\lambda} + \mathcal{O} (\lambda^0)$, very close to the exact value of Eq.~\eqref{eq:2plaq_gap}. 

\begin{figure}
    \centering
    \includegraphics[width=0.57\linewidth]{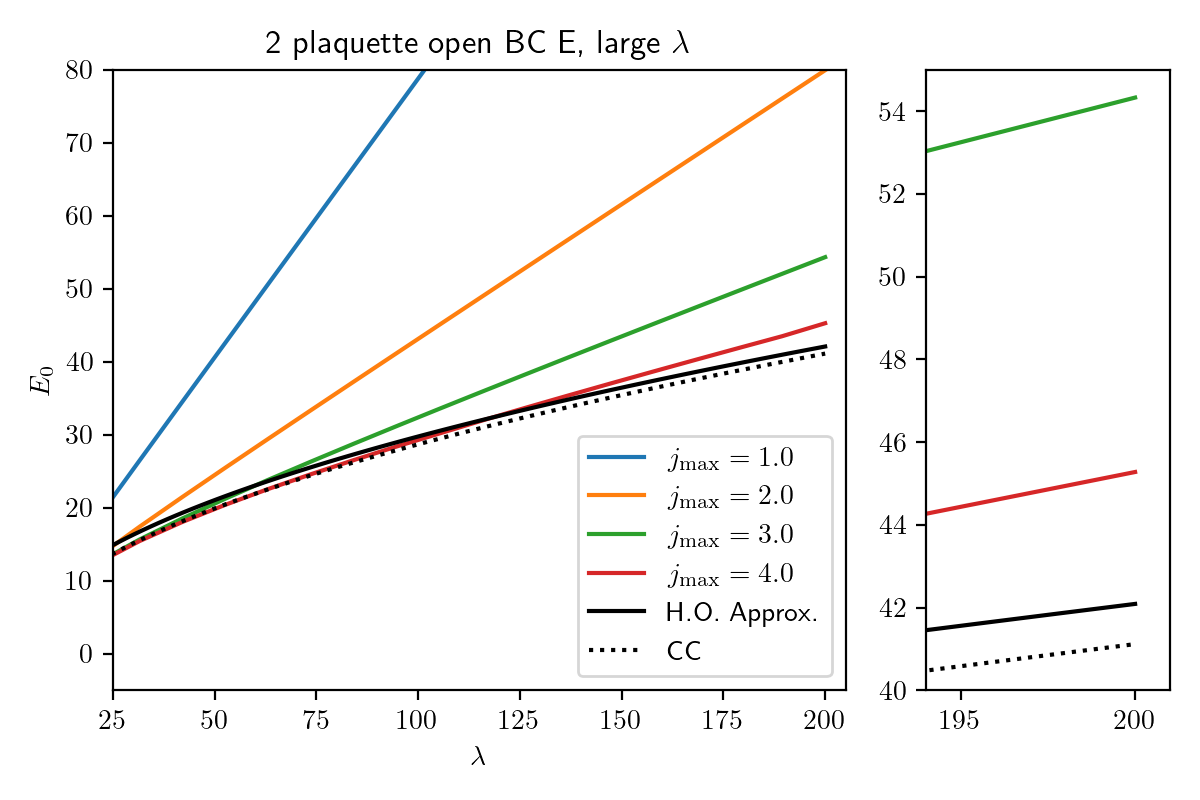}
    \includegraphics[width=0.42\linewidth]{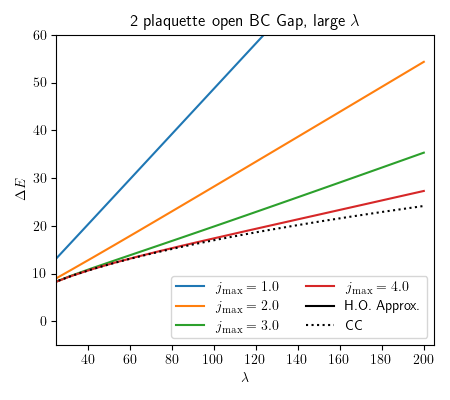}
    \caption{Ground state energy (left panel) and energy gap (right panel) for two plaquettes using the coupled cluster ansatz in the large $\lambda$ limit compared with exact diagonalization with specified truncation and harmonic oscillator (H.O.) approximation.}
    \label{fig:fig2plaqlargelambda}
\end{figure}

The method discussed above generalizes to $N$ plaquettes in a single row with open boundary conditions. The system contains $2N+2$ sites joined by $3N+1$ gauge links. We can use Gauss's law on $2N+1$ sites to fix an equal number of links, leaving $N$ independent $SU(2)$ matrices $X(a)$ ($a=1,2,\dots, N$) that describe the system. These matrices can be chosen conveniently so that the Wilson loops, which enter the magnetic part of the Hamiltonian, have a simple structure
\be 
W[C_a] = \frac{1}{2} \mathrm{Tr} X(a) = x_0 (a) \,.
\ee 
The remaining Gauss's law equation is used to impose zero total angular momentum on the physical states. In the large-$\lambda$ limit, the Hamiltonian reduces to
\be \label{eq:D1} H = \sqrt{\lambda} H^{(0)} + \mathcal{O} (\lambda^0) \ , \ \ H^{(0)} = \frac{1}{2} {\vec{\bm{P}}}^T \cdot H_E \vec{\bm{P}} + \frac{1}{2} {\vec{\bm{X}}}^T \cdot \vec{\bm{X}} \ee
where
\be \label{eq:52a}
  H_E =
 \frac{1}{4} \left[ {\begin{array}{cccccc}
    4 & 2 & 0 & 0 & \cdots & 0\\
    2 & 4 & 2 & 0 & \cdots & \vdots\\
    0 & 2 & 4 & \ddots & \ddots & \vdots\\
    0 & 0 & \ddots & \ddots & 2 & 0\\
    \vdots & \vdots & \ddots & 2 & 4 & 2\\
    0 & 0 & 0 & 0 & 2 & 4\\
  \end{array} } \right] \ , \ \ \vec{\bm{P}} = \left[ {\begin{array}{c}
   \vec{p} (1) \\
   \vec{p} (2) \\
   \vec{p} (3) \\
    \vec{p} (4) \\
    \vdots \\
    \vec{p} (N) \\
  \end{array} } \right]\ , \ \ \vec{\bm{X}} = \left[ {\begin{array}{c}
   \vec{x} (1) \\
   \vec{x} (2) \\
   \vec{x} (3) \\
    \vec{x} (4) \\
    \vdots \\
    \vec{x} (N) \\
  \end{array} } \right]\,,
\ee
and $[x_i (a) , p_j (b)] = i \delta_{ab} \delta_{ij}$. The matrix $H_E$ is a tridiagonal  Toeplitz matrix, and its eigenvalues have a simple closed form, $\lambda_k = 1 - \frac{1}{2} \cos{\frac{k\pi}{N + 1}}$, $k=1,2,\dots, N$. We deduce the normal frequencies of the system
\be \omega_k = \sqrt{\lambda_k \lambda} = \sqrt{\lambda} \sqrt{1 - \frac{1}{2} \cos{\frac{k\pi}{N + 1}}}\,.\ee
Thus, for a system with $N$ plaquettes and open boundary conditions, the ground state energy in the large $\lambda$ limit is
\begin{equation}\label{eq:40}
    E_0(N) = \frac{3}{2}\sum_{k=1}^N \omega_k + \mathcal{O} (\lambda^0) \,.
\end{equation}
Since the lowest frequency is $\omega_1$, the energy of the first excited state is
$E_1(N) \approx 2\omega_1 + E_0(N)$,\footnote{Note that the first excited state of the three-dimensional harmonic oscillator with frequency $\omega_1$ is not allowed due to Gauss's law, hence the energy difference of $2\omega_1$ with the ground state.}
and the gap is
\begin{equation}\label{eq:55g}
    \Delta E (N) = E_1(N)-E_0(N) = 2\sqrt{\lambda}\sqrt{1-\frac{1}{2}\cos{\frac{\pi}{N+1}}} + \mathcal{O} (\lambda^0) \,.
\end{equation}
Note that in the limit $N\to\infty$, the gap approaches a finite value, $\Delta E \to \sqrt{2\lambda}$. For this system, the continuum limit does not exist. This is expected because there is no magnetic field in one spatial dimension.

\begin{figure}
    \centering
    \includegraphics[width=0.47\linewidth]{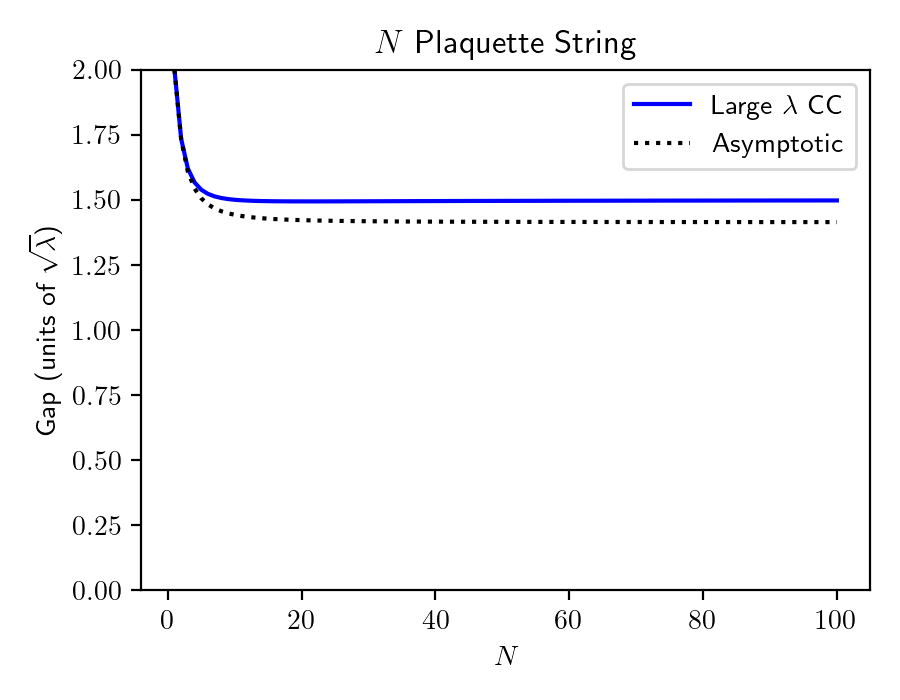}
    \caption{Energy gap for the $N$ plaquette chain in units of $\sqrt{\lambda}$ for the large $\lambda$ coupled cluster (CC) approximation in Eq.~\eqref{eq:55h} (blue solid) and exact value in Eq.~\eqref{eq:55g} (black dotted).  }
    \label{fig:Nplaqgap}
\end{figure}

The system consists of coupled harmonic oscillators in the large $\lambda$ limit. It is convenient to introduce normal modes $\vec{v}_k = {\bm{e}}_k^T \cdot \vec{\bm{X}}$, where $\bm{e}_k$ is the unit eigenvector of $H_E$ corresponding to eigenvalue $\lambda_k$. The ground and first excited states are
\be\label{eq:59psi} \Psi_0^{(0)} \propto e^{-\frac{1}{2} \sum_k \omega_k \vec{v}_k^{\, 2}} \ , \ \ \Psi_1^{(0)} \propto ( \omega_1 \vec{v}_1^{\, 2} - 3) \Psi_0 \,.\ee
To obtain the energy gap with a variational method, consider the coupled cluster ansatz 
\be\label{eq:57a} \psi_0^{(0)}(\alpha) \propto e^{-\frac{\alpha}{2} \sum_k \vec{v}_k^{\, 2}} \ , \ \ \psi_1^{(0)} (\alpha) \propto \left(  \vec{v}_1^{\, 2} - \beta \right) \psi_0 (\alpha)\,, \ee
where $\beta$ is determined by demanding orthogonality of the two wavefunctions
\be \beta =  \frac{3}{2\alpha }\,.\ee
We obtain the following estimate of the ground state energy
\be \epsilon_0^{(0)} (\alpha) = \frac{3N\alpha}{4}  + \frac{3}{4\alpha} \sum_k\lambda_k^2 = \frac{3N}{4} \left( \alpha + \frac{1}{\alpha} \right) \,,\ee
which is minimized for $\alpha = 1$, yielding the estimate $E_0 = \sqrt{\lambda}\epsilon_0^{(0)} (1) = \frac{3N}{2} \sqrt{\lambda}$. It is in good agreement (within 4$\%$) with the exact value in Eq.~\eqref{eq:40} which, for large $N$, is 
\be E_0 (N) \approx \frac{3N}{2}\sqrt{\lambda} \int_0^1 dx \sqrt{1- \frac{1}{2} \cos \pi x}~ \approx 0.98 \frac{3N}{2} \sqrt{\lambda}\,.\ee
After calculating the energy of the state $\ket{\psi_1(\alpha)}$ (Eq.~\eqref{eq:57a}), and setting $\alpha =1$, we deduce the following estimate of the energy gap
\be\label{eq:55h} \Delta E_{\mathrm{est.}} (N) \approx \left( 2 - \frac{1}{2}   \cos \frac{\pi }{N+1} \right) \sqrt{\lambda} \ee
to be compared with Eq.~\eqref{eq:55g}. In Figure \ref{fig:Nplaqgap}, we show that there is good agreement between the asymptotic expression in Eq.~\eqref{eq:55g} and the estimate in Eq.~\eqref{eq:55h} from the coupled cluster ansatz in Eq.~\eqref{eq:57a}. In the large $N$ limit, we obtain the estimate $\Delta E_{\mathrm{est.}} (N) \approx 1.5\sqrt{\lambda}$ which agrees with the actual value $\Delta E = \sqrt{2\lambda} \approx 1.414 \sqrt{\lambda}$ to within $6\%$. 

\section{Two spatial dimensions}
\label{sec:V}

We now discuss the case of two spatial dimensions. A square lattice of $(N+1)\times (N+1)$ sites contains $N^2$ plaquettes and a total of $2N(N+1)$ gauge links. After using Gauss's law at all but one site, we fix $(N+2)N$ links, leaving $N^2$ physical links matching the number of plaquettes. Gauss's law at the remaining site further restricts physical states to the states of zero total angular momentum. For large $\lambda$, the mass gap for a square lattice with $N\times N$ plaquettes is given by \cite{Muller:1983rs}
\be\label{eq:60g} \Delta E_{N\times N} = 2\sqrt{2\lambda}\sin{\frac{\pi}{2(N+1)}} + \mathcal{O} (\lambda^0)\ . \ee
The gap vanishes in the large volume limit ($N \to \infty$) in the strict $\lambda \to \infty$ limit. However, it should be noted that the gap is finite as $N\to\infty$ if we include next-to-leading-order corrections in $1/\lambda$, as discussed in Refs.~\cite{Muller:1983rs, Feynman1981}. 

The simplest case of the square lattice is shown in Fig.~\ref{fig:2DGF}. We fix the gauge links in the maximal tree to identity matrix~\cite{Creutz:1976ch} and set the origin of the tree at the base of links 1 and 4, shown by the red circle in Fig.~\ref{fig:2DGF}. The choice of constructing a maximal tree is not unique and can be selected to result in the simplest description of the system. We choose a maximal tree such that the remaining physical links are:
\begin{figure}[ht!]
\begin{center}
\tikzset{->-/.style={decoration={
  markings,
  mark=at position .5 with {\arrow{>}}},postaction={decorate}}}
\begin{tikzpicture} 
     \draw[->-] (0,0) to (3,0); \draw[->-] (3,0) to (3,3); \draw[blue,thick,dashed][->-] (0,3) to (3,3); \draw[->-] (0,0) to (0,3); \draw[->-] (0,3) to (0,6); \draw[->-] (3,3) to (3,6); \draw[blue,thick,dashed][->-] (0,6) to (3,6) ; \draw[->-] (3,0) to (6,0); \draw[blue,thick,dashed][->-] (3,3) to (6,3); \draw[->-] (6,0) to (6,3); \draw[->-] (6,3) to (6,6); \draw[blue,thick,dashed][->-] (3,6) to (6,6) ;\node[anchor = north] at (1.5,0) {${U(1)} \to \mathbb{I}$}; \node[anchor = north] at (4.5,0) {${U(5)}\to \mathbb{I}$};  \node[anchor = south] at (1.5,3) {${U(3)} \to {X}(1)$}; \node[anchor = south] at (4.5,3) {${U(11)} \to {X}(3)$};  \node[anchor = west] at (-1.7,1.5) {${U(4)}\to \mathbb{I}$}; \node[anchor = west] at (3,1.5) {${U({2})}\to \mathbb{I}$};  \node[anchor = west] at (6,1.5) {${U({6})}\to \mathbb{I}$};
     \node[anchor = west] at (-1.8,4.5) {${U({10})}\to \mathbb{I}$};
     \node[anchor = west] at (3,4.5) {${U({12})}\to \mathbb{I}$};
     \node[anchor = west] at (6,4.5) {${U({7})}\to \mathbb{I}$};
     \node[anchor = south] at (1.5,6) {${U({9})} \to {X}(2)$};
     \node[anchor = south] at (4.5,6) {${U({8})} \to {X}(4)$};
      \fill[red] (0,0) circle (3pt);
\end{tikzpicture}
\end{center}
\caption{Spatial square lattice consisting of $2\times2$ plaquettes that are gauge-fixed by the maximal tree procedure. The blue dashed links are the physical links and solid lines are the tree links fixed to the identity.  
} \label{fig:2DGF}
\end{figure}
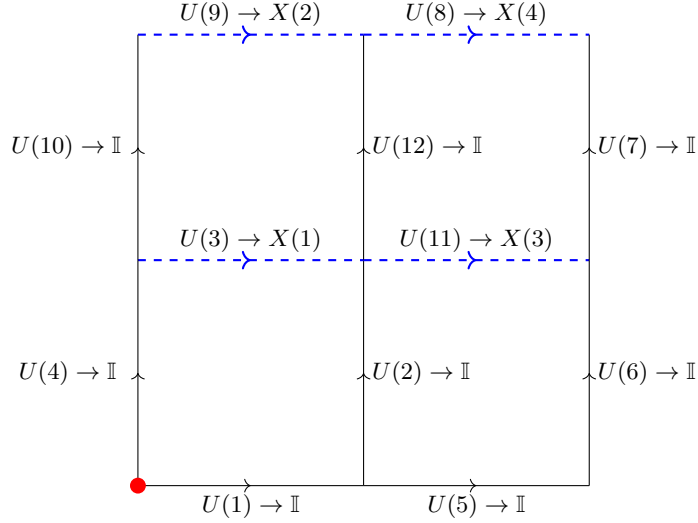
\begin{align}
    {X}(1) &= {U}(4){U}(3){U}^{\dagger}(2){U}^{\dagger}(1), & {X}(2) &= {U}(1) {U}(10) {U}(9) {U}^{\dagger} (12) {U}^{\dagger} (2) {U}^{\dagger} (1) \nonumber\\
    {X}(3) &= {U}(1) {U}(2) {U}(11) {U}^{\dagger} (6) {U}^{\dagger} (5) {U}^{\dagger} (1) & {X}({4}) &= {U}(1) {U}(2) {U}({12}) {U}({8}) {U}^{\dagger} (7) {U}^{\dagger} (6) {U}^{\dagger} (5) {U}^{\dagger} (1)
\end{align}
After changing variables $X(2) \to X(2) X(1)^\dagger$ and $X(4) \to X(4) X^\dagger (3)$,
the Hamiltonian reads
\begin{align}\label{eq:65h}
H &=  \frac{1}{2} \Big( 2\sum_{a=1}^4 (\vec{L}^2(a)+\vec{K}^2(a)) +(\vec{L}(1)+\vec{K}(1))\cdot(\vec{L}(2)-\vec{K}(2)) +(\vec{L}(3)+\vec{K}(3))\cdot (\vec{L}(1)-\vec{K}(1)) \nonumber \\ &+ (\vec{L}(4)+\vec{K}(4))\cdot (\vec{L}(2)-\vec{K}(2)) + (\vec{L}(3)+\vec{K}(3))\cdot (\vec{L}(4)-\vec{K}(4))\Big) + \lambda \Big( 4 - \sum_{a=1}^4 x_0(a) \Big)
\end{align}
In the large $\lambda$ limit, we obtain
\begin{align}
    \vec{x} (1) &= \frac{1}{2} [\vec{u} (1) + \vec{u} (2) - \vec{u} (3) - \vec{u} (4)  ] \nonumber\\
    \vec{x} (2) &= \frac{1}{2} [\vec{u} (9) - \vec{u} (2) + \vec{u} (10) - \vec{u} (12)  ] \\  
    \vec{x} ({3}) &= \frac{1}{2} [\vec{u} (2) - \vec{u} (5) - \vec{u} (6) + \vec{u} (11)  ] \nonumber \\
    \vec{x} ({4}) & = \frac{1}{2} [\vec{u} (12) - \vec{u} (7) + \vec{u} (8) - \vec{u} (11) ] \,.
\end{align}
The Hamiltonian in Eq.~\eqref{eq:65h} reduces to
\be \label{eq:D1a} H = \sqrt{\lambda} H^{(0)} + \mathcal{O} (\lambda^0) \ , \ \ 
 H^{(0)} = \frac{1}{2} {\vec{\bm{P}}}^T \cdot H_E \vec{\bm{P}} + \frac{1}{2} {\vec{\bm{X}}}^T \cdot \vec{\bm{X}} \,,\ee
where
\be\label{eq:65HPX}
  H_E =
 \frac{1}{4}
    \begin{pmatrix}
        4 &- 1 & -1 & 0 \\
        -1 & 4 & 0 & -1\\
        -1 & 0 & 4 & -1 \\
        0 & -1 & -1 & 4
    \end{pmatrix} \ , \ \ \vec{\bm{P}} = \begin{pmatrix}
   \vec{p} (1) \\
   \vec{p} (2) \\
   \vec{p} (3) \\
    \vec{p} (4) 
  \end{pmatrix} \ , \ \ \vec{\bm{X}} = \begin{pmatrix}
   \vec{x} (1) \\
   \vec{x} (2) \\
   \vec{x} (3) \\
    \vec{x} (4)  
  \end{pmatrix}
\ee
and $[x_i (a) , p_j (b)] = i \delta_{ab} \delta_{ij}$.
The eigenvalues of $H_E$ are $\{ \frac{1}{2},1,1, \frac{3}{2} \}$, therefore, the frequencies and corresponding normalized eigenvectors are 
\begin{align}\label{eq:66oe}
    \omega_1 &= \sqrt{\frac{\lambda}{2}} \ , \quad \bm{e}_1 = \frac{1}{2} \begin{pmatrix}
        1, & 1, &
       -1,  & -1
    \end{pmatrix}^T \nonumber\\
    \omega_2 &= \sqrt{\lambda} \ , \quad \bm{e}_2 = \frac{1}{\sqrt{2}}\begin{pmatrix}
        1, & 0, &
       0,  & 1
    \end{pmatrix}^T \nonumber\\
    \omega_3 &= \sqrt{\lambda} \ , \quad \bm{e}_3  = \frac{1}{\sqrt{2}}\begin{pmatrix}
        0, & 1, &
       1,  & 0
    \end{pmatrix}^T \nonumber\\
    \omega_4 &= \sqrt{\frac{3\lambda}{2}} \ , \quad \bm{e}_4 = \frac{1}{2}\begin{pmatrix}
        1, & -1, &
       1,  & -1
    \end{pmatrix}^T
\end{align}
The ground state has energy $E_0 = \frac{3}{2}\sum_i\omega_i + \mathcal{O} (\lambda^0) \approx  5.9\sqrt{\lambda} + \mathcal{O} (\lambda^0)$. The first excited state is obtained by exciting the normal mode with the lowest frequency $\omega_1$, thus the gap is $E_1 - E_0 = 2\omega_1 = \sqrt{2\lambda}$, in agreement with the general expression Eq.~\eqref{eq:60g} for $N=2$ \cite{Muller:1983rs}.  Expressions for the ground and first excited states are given in Eq.~\eqref{eq:59psi} with normal mode coordinates $\vec{v}_k = \bm{e}_k\cdot \vec{\bm{X}}$, where $\bm{X}$ is defined in Eq.~\eqref{eq:65HPX} and the unit vectors $\bm{e}_k$ ($k=1,2,3,4$) are defined in Eq.~\eqref{eq:66oe}.

\begin{figure}
    \centering
    \includegraphics[width=0.47\linewidth]{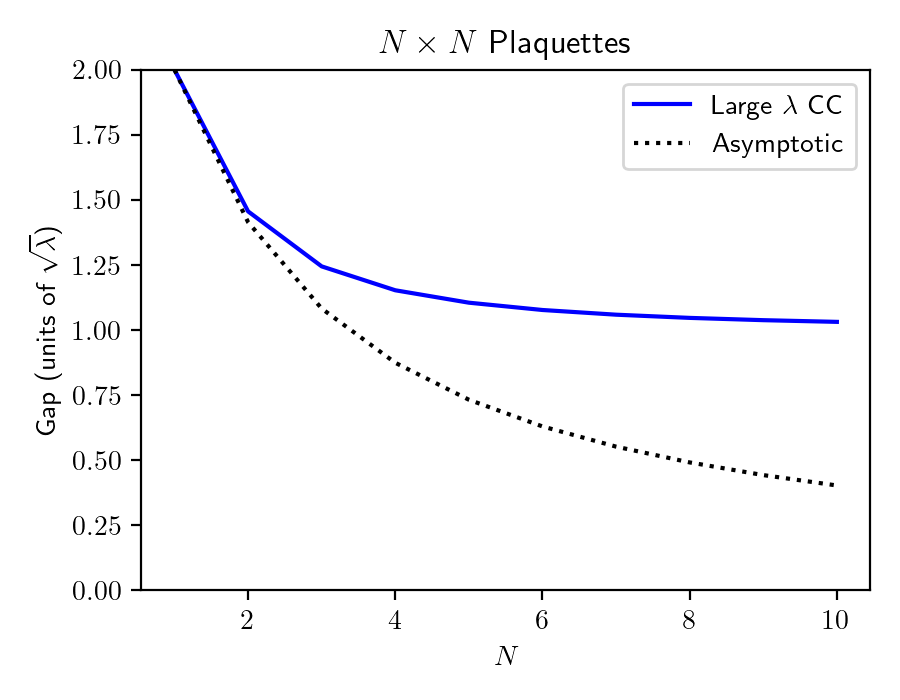}
    \caption{Energy gap for a square lattice with $N\times N$ plaquettes in units of $\sqrt{\lambda}$ in the large $\lambda$ limit using the coupled cluster (CC) approximation in Eq.~\eqref{eq:70de} and the asymptotic expression in Eq.~\eqref{eq:60g}. In the large-$N$ limit, the asymptotic value is 0 while the CC gives 1 indicating a breakdown of the ansatz. One way to improve this is to add higher-order terms in the ansatz similar to Eq.~\eqref{eq:68psiN}}
    \label{fig:Nbyplaqgap}
\end{figure}

To obtain the energy gap with a variational method, consider the coupled cluster ansatz in Eq.~\eqref{eq:57a}, which is valid in the large $\lambda$ limit.
For the ground state energy, we obtain
\be \epsilon_0^{(0)} (\alpha) =   3 \left( \alpha + \frac{1}{\alpha} \right) \,,\ee
which is minimized for $\alpha =1$. We deduce the estimate of the ground state energy $E_0 = 6\sqrt{\lambda} + \mathcal{O} (\lambda^0)$ which is close to the exact value. Similarly, we obtain the first excited energy $E_0 = 7.5\sqrt{\lambda} + \mathcal{O} (\lambda^0)$. We deduce the estimate of the gap, $\Delta E = 1.5\sqrt{\lambda} + \mathcal{O} (\lambda^0)$, which is a good approximation to the exact value. 

Away from the large-$\lambda$ limit, we define the following extension of the coupled cluster ansatz
\be\label{eq:68psi} \psi_0 (\alpha) \propto e^{\alpha \sum_a x_0(a)} \ , \ \ \psi_1 (\alpha) \propto \left( \sum_{ab} A_{ab} \bm{x} (a) \cdot \bm{x} (b) - \beta \right) \psi_0 (\alpha) \,,\ee
where $\beta$ is determined by demanding orthogonality of the two states and the coefficients $A_{ab}$ are chosen so that this ansatz matches Eq.~\eqref{eq:57a} in the large-$\lambda$ limit. Note that in that limit, $\bm{x} (a) \cdot \bm{x} (b) \approx -\frac{1}{2} (\vec{x} (a) - \vec{x} (b))^2$. Using Eq.~\eqref{eq:66oe}, we have
\be \vec{v}_1^{\,2} = (\bm{e}_1\cdot \vec{\bm{X}})^2 = \frac{1}{2} (\vec{x} (1) + \vec{x} (2) - \vec{x} (3) - \vec{x} (4))^2 \,. \ee
Therefore, $\vec{v}_1^{\, 2}$ is recovered from
\be \bm{x} (1) \cdot \bm{x} (2) - \bm{x} (1) \cdot \bm{x} (3) - \bm{x} (1) \cdot \bm{x} (4) - \bm{x} (2) \cdot \bm{x} (3) - \bm{x} (2) \cdot \bm{x} (4) + \bm{x} (3) \cdot \bm{x} (4) \ee
in the large-$\lambda$ limit, which fixes $A_{12} = -A_{13} = - A_{14} = -A_{23} = - A_{24} = A_{34} = \frac{1}{2}$. The ansatz Eq.~\eqref{eq:68psi} can also be expressed in terms of Wilson loops. The calculation of energy levels using the Hamiltonian in Eq.~\eqref{eq:65h} proceeds as in the previous cases.

Larger lattice sizes can be tackled by extending the above approach.
For $N\times N$ plaquettes, using the ansatz Eq.~\eqref{eq:57a}, we obtain the following lowest energy levels in the large-$\lambda$ limit
\begin{align} \label{eq:68nxn}
\epsilon_0^{(0)} (\alpha) &= \frac{3N^2}{4} \left( \alpha + \frac{1}{\alpha} \right) \,,\nonumber\\
    \epsilon_1^{(0)}(\alpha) &=  \left( \alpha\lambda_1 + \frac{1}{\alpha} \right) \frac{  5 \alpha ^{2}+4 \alpha -1  }{5 \alpha ^{2}-2 \alpha +5  }    + \epsilon_0^{(0)} (\alpha)\,,
\end{align}
where $\lambda_1$ is the lowest eigenvalue
\be \lambda_1 = 2\sin^2 \frac{\pi}{2(N+1)}\,. \ee
The ground state energy is minimized for $\alpha =1$. We obtain the estimate $E_0 = \frac{3N}{2} \sqrt{\lambda} + \mathcal{O} (\lambda^0)$. For the gap, setting $\alpha =1$, we obtain the estimate
\be\label{eq:70de} \Delta E = \left( 1+ 2\sin^2 \frac{\pi}{2(N+1)} \right) \sqrt{\lambda} + \mathcal{O} (\lambda^0)  \ee
A comparison of this estimate with the asymptotic expression in Eq.~\eqref{eq:60g} is shown in Figure \ref{fig:Nbyplaqgap}. In the large volume limit, the coupled cluster estimate in Eq.~\eqref{eq:70de} approaches $\sqrt{\lambda}$ whereas the exact value vanishes. The coupled cluster ansatz of  Eq.~\eqref{eq:57a} fails. To improve the performance in the important large volume (continuum) limit, we introduce a modified ansatz by adding a two-body term to the ansatz of Eq.~\eqref{eq:68psi} that reproduces the exact expression Eq.~\eqref{eq:59psi} in the large $\lambda$ limit
\be\label{eq:68psiN} \psi_0 (\alpha) \propto e^{\alpha \left( \sum_a x_0(a) +  \sum_{ab} B_{ab} \bm{x} (a) \cdot \bm{x} (b) \right)} \ , \ \ \psi_1 (\alpha) \propto \left( \sum_{ab} A_{ab} \bm{x} (a) \cdot \bm{x} (b) - \beta \right) \psi_0 (\alpha) \,.\ee
Here, $\alpha$ is a variational parameter and $\beta$ is determined by demanding orthogonality of the two states. The coefficients $A_{ab}, B_{ab}$ are chosen so that this ansatz matches the form of the expression in Eq.~\eqref{eq:59psi} in the large-$\lambda$ limit. This ansatz reproduces the asymptotic expression in Eq.~\eqref{eq:60g} for the gap and can capture more accurately the physics in the continuum limit.

\section{Quantum computation with qumodes}\label{sec:VI} 

We consider two approaches to the quantum computation (QC) with qumodes for $SU(2)$ lattice gauge theory. For the first one, we use quadruplets of qumodes to represent each physical link and the second one relies on qumode triplets. In both approaches, the number of required qumodes grows \textit{linearly} with the spatial volume of the lattice. 
 
\subsection{QC with qumode quadruplets}

An $SU(2)$ matrix can be written as $U = u_0 \mathbb{I} + \vec{u}\cdot \vec{\sigma}$, where $u_0^2 + \vec{u}^{\,2} = 1$. To represent this with qumodes, we introduce a quadruplet of qumodes with quadratures $\bm{q} = (q_0,q_1,q_2,q_3)$ and $\bm{p} = (p_0,p_1,p_2,p_3)$, obeying the commutation rules $[q_\mu , p_\nu] =i\delta_{\mu\nu}$ ($\mu,\nu = 0,1,2,3$). However, these quadratures take values on the entire real axis. To restrict them to the unit sphere, we consider wave functions $\psi (\bm{q})$, which only have support close to the unit sphere ($\bm{q}^2 \approx 1$). We impose this constraint by including a factor of the form
\be\label{eq:85} e^{-\frac{\Lambda^2}{2} (\bm{q}^2 -1)^2} \,,\ee
which reduces to a Dirac $\delta$-function $\delta (\bm{q}^2 -1)$ in the limit $\Lambda\to\infty$. We will engineer this factor with a quantum circuit similar to the one introduced in \cite{Jha:2023ecu} for the simpler $O(3)$ model in 1+1 dimensions. 

\subsubsection{Single plaquette}

\begin{figure}
      \centering
      \includegraphics[width=0.57\linewidth]{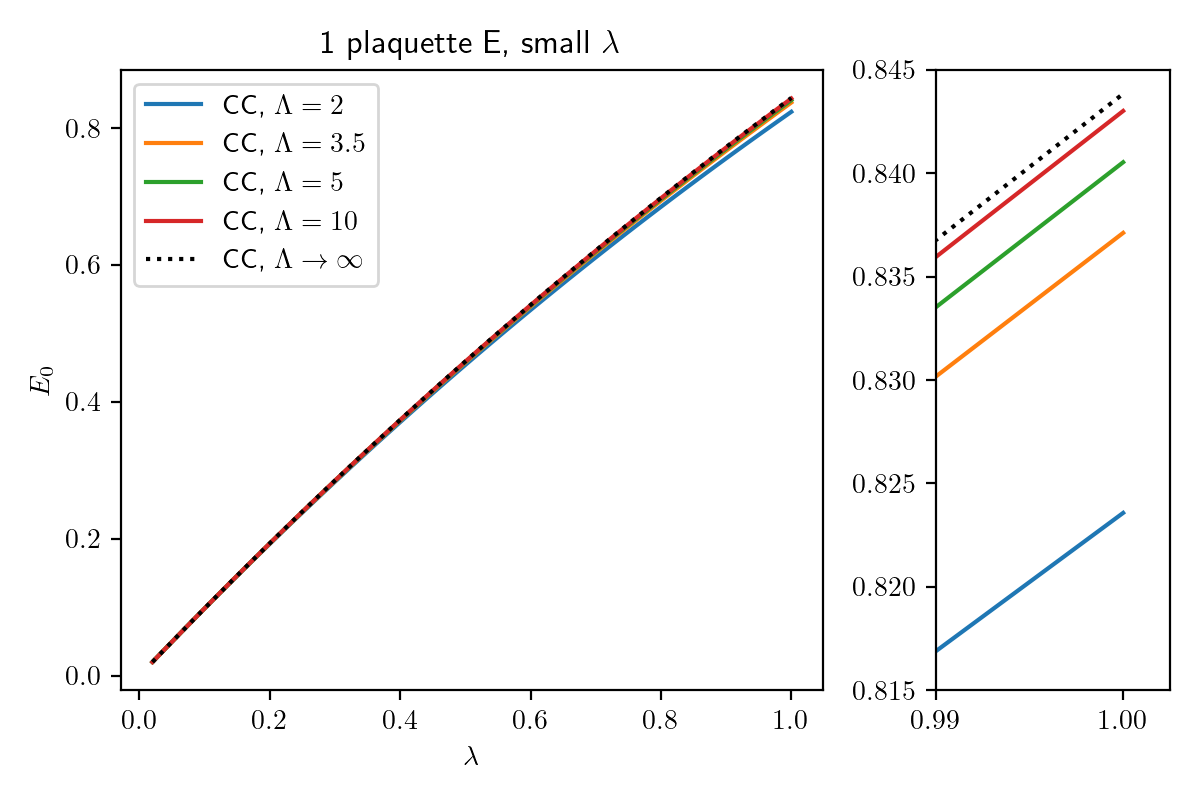}
      \includegraphics[width=0.42\linewidth]{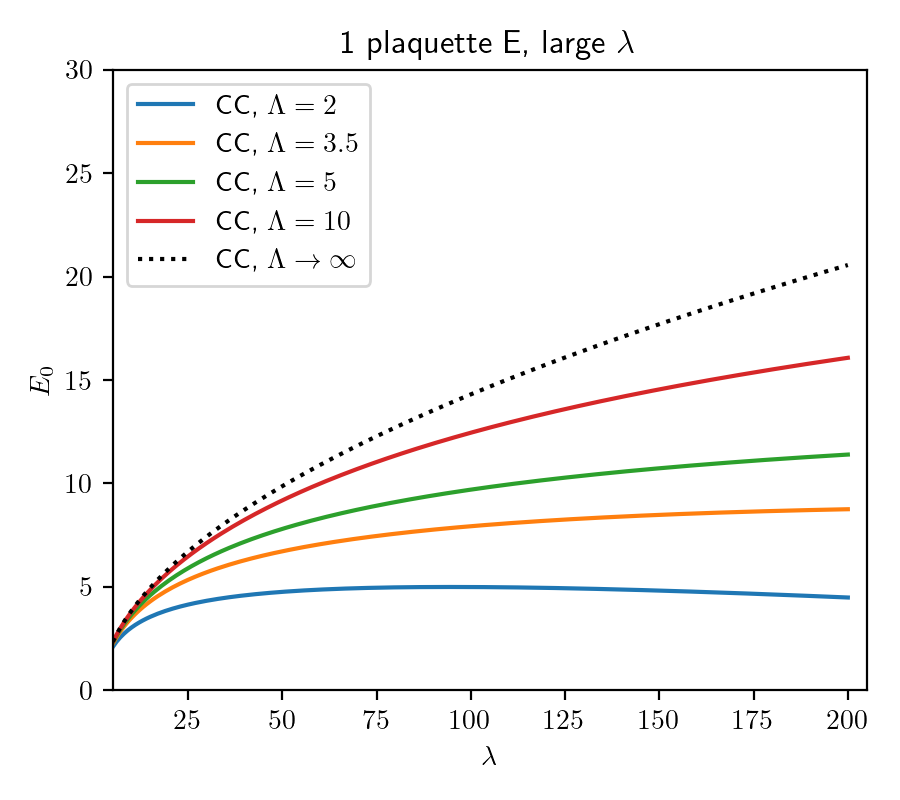}
      \caption{Ground state energy of a single plaquette for small (left) and large (right) $\lambda$ using the coupled cluster (CC) ansatz \eqref{eq:87} for various values of the cutoff $\Lambda$. }
      \label{fig:fig10}
  \end{figure}

Recall that for a single plaquette, after fixing the gauge, the Hamiltonian is given by  Eq.~\eqref{eq:H1pl}. For QC, we will use a single qumode quadruplet with quadratures $\bm{q} = (q_0,\vec{q})$ and $\bm{p} = (p_0,\vec{p})$, and work with the Hamiltonian
\begin{equation}\label{eq:85H}
H^{\text{QC}} =\frac{1}{2}(\vec{L}^2 + \vec{K}^2) + \lambda (1-q_0) \,,
\end{equation}
where $\vec{L} = \vec{q}\times \vec{p}$ and $\vec{K} = \vec{q} p_0 - q_0 \vec{p}$. 

To compute the ground state energy, we introduce the coupled cluster ansatz. In the $q$-quadrature representation, it reads
\begin{equation}\label{eq:87}
    \braket{\bm{q}|\psi_{0}^{\text{QC}} (\alpha)} \propto \Phi(\alpha;\bm{q}) \ , \ \  \Phi(\alpha;\bm{q}) = e^{\alpha q_0}\ e^{-\frac{\Lambda^2}{2}(\bm{q}^2-1)^2}\,,
\end{equation}
where we defined the function $\Phi$ for later use, to be compared with the ansatz in Eq.~\eqref{eq:23} used in the classical calculation. Its energy is found to be
\be \epsilon_0^{\text{QC}} (\alpha) = \bra{\psi_0^{\text{QC}} (\alpha)} H^{\text{QC}} \ket{\psi_0^{\text{QC}} (\alpha)} = \epsilon_0 (\alpha) + \frac{\alpha (3\alpha - 4\lambda)}{16\Lambda^2} + \mathcal{O} (\Lambda^{-4}) \ee
reproducing the energy in Eq.~\eqref{eq:24} of the trial state in the classical calculation at leading order in $1/\Lambda$, as desired. The energy of the ground state is plotted for various values of the cutoff parameter $\Lambda$ in Figure \ref{fig:fig10}. It approaches the classical result in the limit $\Lambda \to \infty$. Notice that for finite $\Lambda$ we obtain values below the ground state energy of the Hamiltonian \eqref{eq:H1pl}. This is possible because we have replaced the Hamiltonian with $H^{QC}$ (Eq.\ \eqref{eq:85H}) for quantum computations, which is unbounded from below.

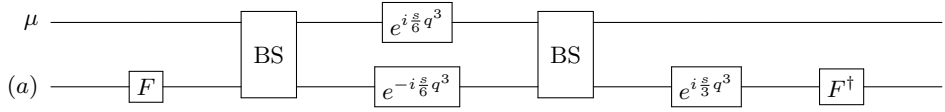
\begin{figure}[ht!]
    \centering
\[\Qcircuit @C=3.2em @R=1em { \lstick{ \mu}  &\qw & \multigate{1}{\text{BS}} & \gate{e^{i \frac{s}{6} q^3}} & \multigate{1}{\text{BS}} & \qw &\qw  &\qw \\
   \lstick{(a)} & \gate{F} &\ghost{\text{BS}} &\gate{e^{-i \frac{s}{6} q^3}} & \ghost{\text{BS}} &\gate{e^{i \frac{s}{3} q^3}} & \gate{F^\dagger}   &\qw    } \]
    \caption{Quantum circuit that generates a factor of the form $e^{-is {q}_\mu^2 {p}^{(a)}}$, Eq.~\eqref{eq:79}, for the non-linear controlled gate Eq.~\eqref{eq:79D}.}
    \label{fig:QC2C}
\end{figure}

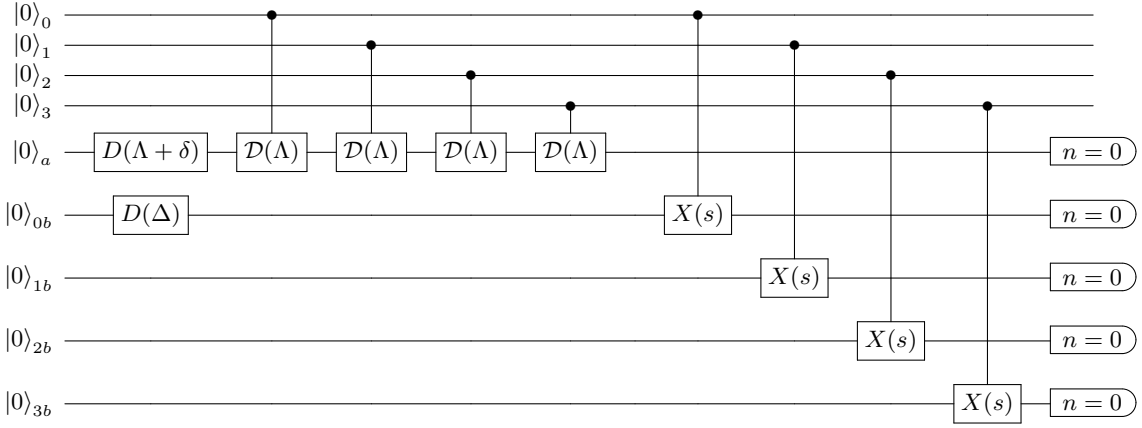
\begin{figure}[ht!]
    \centering
\[\Qcircuit @C=1.2em @R=1em { \lstick{\ket{0}_0} &\qw & \ctrl{4} & \qw & \qw & \qw & \qw & \ctrl{5} & \qw & \qw & \qw  &\qw \\
 \lstick{\ket{0}_1} &\qw &\qw & \ctrl{3} &\qw & \qw  &\qw & \qw& \ctrl{5} & \qw &   \qw  &\qw  \\  \lstick{\ket{0}_2} &\qw& \qw & \qw & \ctrl{2} & \qw & \qw & \qw & \qw & \ctrl{5} & \qw   & \qw  \\ \lstick{\ket{0}_3} &\qw &\qw & \qw &\qw & \ctrl{1}  &\qw & \qw& \qw & \qw &   \ctrl{5}   & \qw  \\  \lstick{\ket{0}_{a}}  & \gate{D(\Lambda + \delta)}  & \gate{\mathcal{D} (\Lambda)} & \gate{\mathcal{D} (\Lambda)} & \gate{\mathcal{D} (\Lambda)} & \gate{\mathcal{D} (\Lambda)} &  \qw & \qw & \qw & \qw  &\qw  &\measureD{n=0} \\ \lstick{\ket{0}_{0b}} &\gate{D(\Delta)}  &\qw & \qw &\qw & \qw  &\qw & \gate{X(s)}& \qw & \qw &   \qw    & \measureD{n=0} \\  \lstick{\ket{0}_{1b}} &\qw & \qw & \qw & \qw & \qw & \qw & \qw & \gate{X(s)}  &\qw  & \qw  & \measureD{n=0} \\
\lstick{\ket{0}_{2b}} &\qw & \qw &\qw & \qw & \qw & \qw & \qw & \qw & \gate{X(s)}   & \qw &\measureD{n=0} \\ \lstick{\ket{0}_{3b}} &\qw & \qw   & \qw  &\qw & \qw & \qw & \qw & \qw & \qw   &\gate{X(s)}   &\measureD{n=0}   } \]
    \caption{Quantum circuit that generates the state in Eq.~\eqref{eq:95psi}. It implements the trial state in Eq.~\eqref{eq:87} with adjustable parameter $s$, $\delta =  \frac{1+ s^2}{2\Lambda} $, and $\Delta = \frac{\alpha}{s}$.}
    \label{fig:QC0}
\end{figure}

To engineer the trial state in Eq.~\eqref{eq:87}, we prepare the system of four qumodes in the vacuum state, $\ket{\bm{0}} = \ket{0}^{\otimes 4}$. We attach an ancilla qumode of quadratures $(q^{(a)}, p^{(a)})$ and entangle it with the qumodes of the system by applying the unitaries,
\begin{equation}\label{eq:79D}
    \mathcal{D} (\Lambda) = e^{-i\sqrt{2}\Lambda \bm{q}^2 {p}^{(a)}} \ , \qquad {D}_a(\Lambda + \delta)  =  e^{i\sqrt{2}(\Lambda +\delta) {p}^{(a)}}
\end{equation}
where $D$ is a displacement operator acting on the ancilla qumode and $\mathcal{D}$ acts as a conditional displacement on the ancilla controlled by the four qumodes of the system. This \textit{non-linear (quadratic) two-qumode control} can be implemented with non-Gaussian cubic phase gates using
\be\label{eq:79} e^{-is {q}_\mu^2 {p}^{(a)}} = F_a^\dagger \cdot U_{BS} \cdot e^{-i\frac{s}{6} q_\mu^3} \cdot e^{i\frac{s}{6} [q^{(a)}]^3} \cdot U_{BS}^\dagger \cdot e^{i\frac{s}{3} [q^{(a)}]^3} \cdot F_a\,,\ee
where $F= e^{i\frac{\pi}{4} (p^2 + q^2)}$ is the Fourier transform operator and $U_{BS}$ is the beam-splitter gate defined as $e^{-\frac{i}{4} (q_1p_2 - p_1q_2)}$ for qumodes with quadratures $(q_i,p_i)$ ($i=1,2$). 
The corresponding quantum circuit is shown in Figure~\ref{fig:QC2C}.
Then we measure the ancilla qumode projecting it onto the zero photon state. We obtain the state
\be {}_a\bra{0} \mathcal{D} (\Lambda) D_a (\Lambda + \delta) \ket{\bm{0}} \ket{0}_a \propto \int d^4 q\, e^{-\frac{1}{2}(\Lambda\bm{q}^2 -\Lambda - \delta)^2} e^{-\frac{1}{2} \bm{q}^2}\ket{\bm{q}}.\ee
Next, we attach a quadruplet of ancilla qumodes of quadratures $(\bm{q}_{b} , \bm{p}_{b})$ initialized in the ground state, $\ket{\bm{0}}_{b}$. We entangle each ancilla qumode with a qumode in the system by applying four controlled-X (CX) gates, collectively given by
\be 
\bm{U}^{\text{(CX)}}(s) = e^{-is \bm{q} \cdot \bm{p}^{(b)}} \,,
\label{eq:91} 
\ee
and displace the qumode 0 in the ancilla quadruplet by acting with the displacement operator ${D}_{0b} (\Delta) = e^{i\sqrt{2} \Delta p_{0}^{(b)}}$. Finally, we measure the ancilla qumodes projecting them onto their respective zero photon states, $\ket{\bm{0}}_{b}$. We arrive at the state
\begin{equation}\label{eq:95psi}
   \ket{\Psi_0 (s,\delta, \Delta)} \propto {}_{b}\bra{\bm{0}}{}_a\bra{0} {D}_{0b} (\Delta) \bm{U}^{(CX)} (s)\mathcal{D} (\Lambda) {D}_a (\Lambda + \delta) \ket{\bm{0}} \ket{0}_a \ket{\bm{0}}_{b} \propto    \int d^4{q} \  e^{s\Delta q_{0}} \  e^{-\frac{\Lambda^2}{2}(\bm{q}^2-1)^2} e^{ \frac{2\Lambda\delta -1- s^2  }{2} \bm{q}^2} \ket{\bm{q}}
\end{equation}
The parameter $s$ is arbitrary and can be chosen randomly. By choosing $\delta = \frac{1+ s^2}{2\Lambda} $ and $\Delta = \frac{\alpha}{s}$, we obtain the desired state in Eq.~\eqref{eq:87}. The quantum circuit that generates this state is shown in Figure \ref{fig:QC0}.

\begin{figure}
    \centering
    \includegraphics[width=0.57\linewidth]{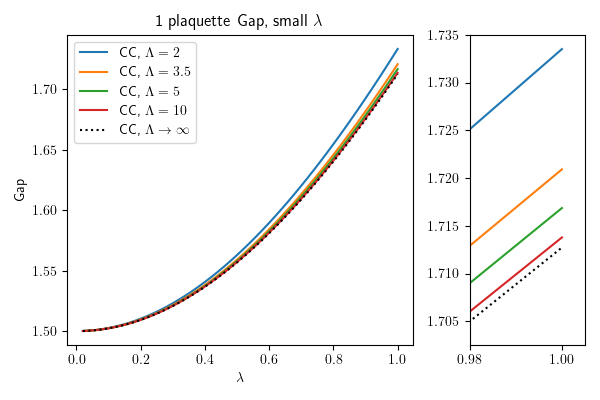}
    \includegraphics[width=0.42\linewidth]{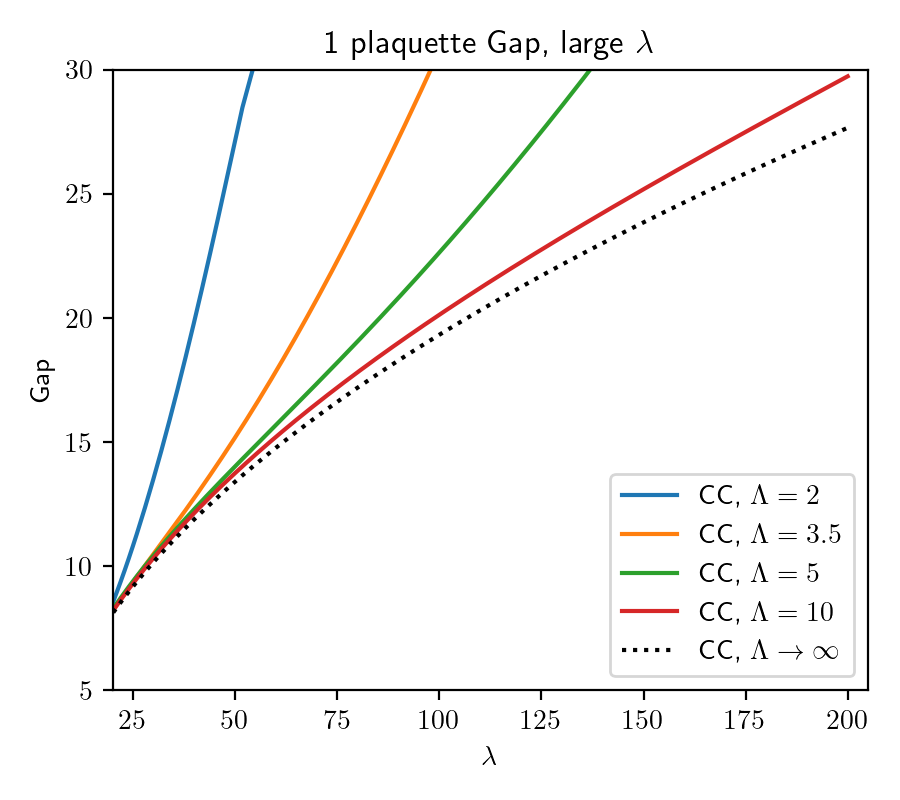}
    \caption{Energy gap of a single plaquette for small (left) and large (right) $\lambda$ using the coupled cluster (CC) ansatz \eqref{eq:87} for various values of the cutoff $\Lambda$. }
    \label{fig:fig10b}
\end{figure}

Similarly, for the first excited state, we adopt the ansatz
\begin{equation}\label{eq:89}
   \ket{\psi_{1}^{\text{QC}} (\alpha)} \propto (q_0 - \beta(\alpha)) \ket{\psi_{0}^{\text{QC}} (\alpha)}\,, \qquad \beta = \frac{I_2(2\alpha)}{I_1(2\alpha)} + \frac{\alpha}{4\Lambda^2} + \mathcal{O} (\frac{1}{\Lambda^4})\,,
\end{equation}
where $\beta$ is chosen by demanding orthogonality, $\braket{\psi_0^{\text{QC}}| \psi_1^{\text{QC}}} =0$,
to be compared with the classical state in Eq.~\eqref{eq:25a}. We recover the expression for the energy in Eq.~\eqref{eq:25e} at leading order in $1/\Lambda$,
\be \epsilon_1^{\text{QC}} (\alpha) = \bra{\psi_1^{\text{QC}} (\alpha)} H^{\text{QC}} \ket{\psi_1^{\text{QC}} (\alpha)} = \epsilon_1 (\alpha) + \mathcal{O} ( \frac{1}{\Lambda^2} )\,. \ee
Figure \ref{fig:fig10b} shows results of the quantum computation for the energy gap ($\Delta E = E_1 - E_0$) and various values of the cutoff parameter $\Lambda$. It approaches the classical result in the limit $\Lambda\to\infty$.

To engineer the trial state in Eq.~\eqref{eq:89}, we will use the same circuit as the one used for the trial state in Eq.~\eqref{eq:87}, shown in Figure \ref{fig:QC0}, except that on the $\mu =0$ qumode of the ancilla quadruplet, we will measure the photon number projecting it onto the single photon number state ($n=1$). We arrive at the state
\begin{equation}\label{eq:93}
 \ket{\Psi_1 (s,\delta, \Delta)} \propto  \int d^4{q}   \left( q_0 - \frac{\Delta}{s} \right)  e^{s\Delta q_{0}} \  e^{-\frac{\Lambda^2}{2}(\bm{q}^2-1)^2} e^{ \frac{2\Lambda\delta -1- s^2  }{2} \bm{q}^2} \ket{\bm{q}}\,.
\end{equation}
By choosing $s = \sqrt{\frac{\alpha}{\beta}}$, $\Delta = \sqrt{\alpha\beta}$, and $\delta = \frac{\alpha + \beta}{2\beta\Lambda}$, we obtain the desired state in Eq.~\eqref{eq:89}. The quantum circuit that generates this state is shown in Figure \ref{fig:QC1}.

\begin{figure}[ht!]
    \centering
\[\Qcircuit @C=1.2em @R=1em { \lstick{\ket{0}_0} &\qw & \ctrl{4} & \qw & \qw & \qw & \qw & \ctrl{5} & \qw & \qw & \qw  &\qw \\
 \lstick{\ket{0}_1} &\qw &\qw & \ctrl{3} &\qw & \qw  &\qw & \qw& \ctrl{5} & \qw &   \qw  &\qw  \\  \lstick{\ket{0}_2} &\qw& \qw & \qw & \ctrl{2} & \qw & \qw & \qw & \qw & \ctrl{5} & \qw   & \qw  \\ \lstick{\ket{0}_3} &\qw &\qw & \qw &\qw & \ctrl{1}  &\qw & \qw& \qw & \qw &   \ctrl{5}   & \qw  \\  \lstick{\ket{0}_{a}}  & \gate{D(\Lambda + \delta)}  & \gate{\mathcal{D} (\Lambda)} & \gate{\mathcal{D} (\Lambda)} & \gate{\mathcal{D} (\Lambda)} & \gate{\mathcal{D} (\Lambda)} &  \qw & \qw & \qw & \qw  &\qw  &\measureD{n=0} \\ \lstick{\ket{0}_{0b}} &\gate{D(\Delta)}  &\qw & \qw &\qw & \qw  &\qw & \gate{X(\gamma)}& \qw & \qw &   \qw    & \measureD{n=1} \\  \lstick{\ket{0}_{1b}} &\qw & \qw & \qw & \qw & \qw & \qw & \qw & \gate{X(\gamma)}  &\qw  & \qw  & \measureD{n=0} \\
\lstick{\ket{0}_{2b}} &\qw & \qw &\qw & \qw & \qw & \qw & \qw & \qw & \gate{X(\gamma)}   & \qw &\measureD{n=0} \\ \lstick{\ket{0}_{3b}} &\qw & \qw   & \qw  &\qw & \qw & \qw & \qw & \qw & \qw   &\gate{X(\gamma)}   &\measureD{n=0}   } \]
    \caption{Quantum circuit that generates the state Eq.~\eqref{eq:93}. It implements the trial state Eq.~\eqref{eq:89} with $s = \sqrt{\frac{\alpha}{\beta}}$, $\Delta = \sqrt{\alpha\beta}$, and $\delta = \frac{\alpha + \beta}{2\beta\Lambda}$.}
    \label{fig:QC1}
\end{figure}
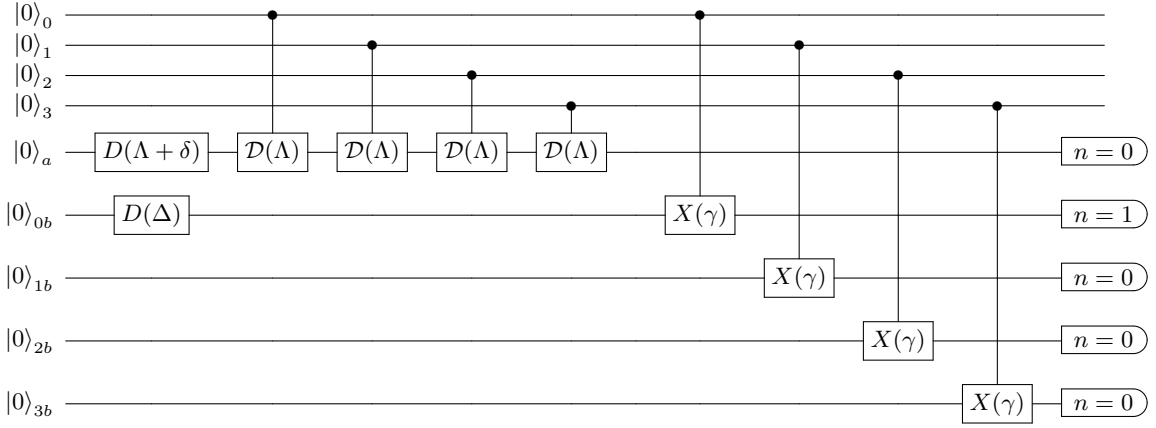


\begin{figure}[ht!]
    \centering
\[ \ \ \ \ \ \ \ \ \  \ \Qcircuit @C=1.7em @R=1em {    \lstick{} & \push{0\ \ \ } & \gate{Z( \frac{\lambda\Delta t}{\sqrt{2}})}& \multigate{1}{e^{-i\frac{\Delta t}{2} J^2}}     &\qw  & \qw & \qw& \link{1}{-1}  &\qw & \multigate{1}{e^{-i\frac{\Delta t}{2} J^2}} &\qw &\link{1}{-1} &\qw &\qw \\
\lstick{} & \push{1\ \ \ } &\qw &\ghost{e^{-i\frac{\Delta t}{2} J^2}}  & \multigate{1}{e^{-i\frac{\Delta t}{2} J^2}} & \qw& \link{1}{-1} & \link{-1}{-1} & \multigate{1}{e^{-i\frac{\Delta t}{2} J^2}} & \ghost{e^{-i\frac{\Delta t}{2} J^2}} &\qw  &\link{-1}{-1} &\link{1}{-1}&\qw \\
\lstick{} & \push{2\ \ \ } &\qw &\multigate{1}{e^{-i\frac{\Delta t}{2} J^2}} &\ghost{e^{-i\frac{\Delta t}{2} J^2}}  & \qw  &\link{-1}{-1}  &\link{1}{-1}   & \ghost{e^{-i\frac{\Delta t}{2} J^2}} &\multigate{1}{e^{-i\frac{\Delta t}{2} J^2}} &\qw &\link{1}{-1} &\link{-1}{-1} &\qw  \\
\lstick{} & \push{3\ \ \ } &\qw & \ghost{e^{-i\frac{\Delta t}{2} J^2}} &\qw & \qw &\qw  &\link{-1}{-1}  &\qw  & \ghost{e^{-i\frac{\Delta t}{2} J^2}} &\qw &\link{-1}{-1} &\qw &\qw } \]
\\
\[\Qcircuit @C=1.7em @R=1em {    \lstick{} & \push{\mu\ \ } & \gate{F} & \multigate{1}{\text{BS}}     &\gate{\mathcal{K} (-\frac{\Delta t}{2})}  & \multigate{1}{\mathcal{K}'({\Delta t})} & \qw & \link{1}{-1} & \multigate{1}{\text{BS}} &\qw&\link{1}{-1}&\gate{F^\dagger}&\qw\\
\lstick{} & \push{\nu\ \ } &\qw &\ghost{\text{BS}}  & \gate{\mathcal{K} (-\frac{\Delta t}{2})} & \ghost{\mathcal{K}'({\Delta t})} &\qw & \link{-1}{-1} & \ghost{\text{BS}}  &\qw&\link{-1}{-1}&\qw &\qw\inputgroupv{1}{2}{.75em}{.75em}{e^{-i\frac{\Delta t}{2} J_{\mu\nu}^2}\ \ \ \ \ \ \ \ \ } }  \]
\caption{Quantum circuit for the time evolution operator in Eq.~\eqref{eq:95ev} of the Hamiltonian in Eq.~\eqref{eq:85H} for a small time step $\Delta t$.}
    \label{fig:evo}
\end{figure}
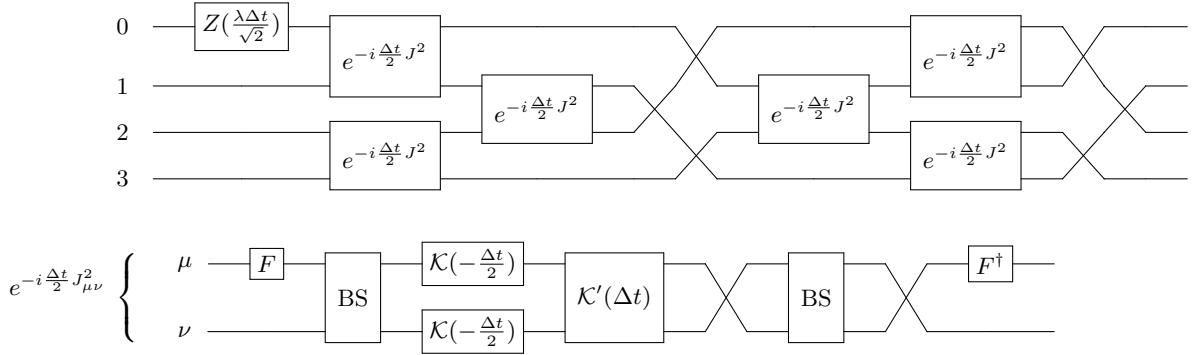

Next, we discuss the time evolution of the system. It can be implemented as a series of small time intervals, $\Delta t$, which compels us to consider the time evolution operator $e^{-i\Delta t H^{\text{QC}}}$,  where $H^{\text{QC}}$ is given by Eq.~\eqref{eq:85H},
\be\label{eq:95ev} e^{-i\Delta t H^{\text{QC}}} \approx e^{i \lambda \Delta t q_0} \prod_{i=1}^3  e^{-i\frac{\Delta t}{2} {L}_i^2} e^{-i\frac{\Delta t}{2} {K}_i^2} \,.\ee
Here we omitted an irrelevant phase $e^{-i\lambda\Delta t}$. The first factor in Eq.~\eqref{eq:95ev} is a displacement operator that shifts $p_0$ by $\frac{\lambda \Delta t}{\sqrt{2}}$,
\be Z_0(s) = e^{i\sqrt{2} s q_0} \ , \ \ s = \frac{\lambda \Delta t}{\sqrt{2}} \ee
Each of the other six factors in Eq.~\eqref{eq:95ev} is of the form $e^{is J_{\mu\nu}^2}$, where $s = \frac{\Delta t}{2}$ and $J_{\mu\nu} = -i (q_\mu p_\nu - q_\nu p_\mu)$  ($\mu,\nu = 0,1,2,3$). They can be implemented with non-Gaussian Kerr gates $\mathcal{K}$ and cross-Kerr gates $\mathcal{K}'$, 
\be \mathcal{K}_\mu(s) = e^{isN_\mu^2} \ , \ \ \mathcal{K}'_{\mu\nu} (s) = e^{isN_\mu N_\nu} \ , \ \ N_\mu = \frac{1}{2} (p_\mu^2 + q_\mu^2)\ . \ee
To this end, note that
\be\label{eq:102J} J_{\mu\nu} = F_\mu^\dagger\cdot U_{BS}\cdot (N_\mu - N_\nu)\cdot U_{BS} \cdot F_\mu \,.\ee
Therefore,
\be e^{is J_{\mu\nu}^2} = F_\mu^\dagger\cdot U_{BS}^\dagger \cdot \mathcal{K}_\mu (s) \cdot \mathcal{K}_\nu (s) \cdot \mathcal{K}_{\mu\nu}' (-2s)  \cdot U_{BS} \cdot F_\mu\,, \ee
where $F_\mu = e^{i\frac{\pi}{2} N_\mu}$ is the Fourier transform operator.
The quantum circuit that implements the evolution operator in Eq.~\eqref{eq:95ev} for a small time interval $\Delta t$ is shown in Figure \ref{fig:evo}.

Matrix elements of the evolution operator are of particular interest. The transition amplitude
\be G(t) = \braket{\mathrm{out} \vert e^{-it H^{\text{QC}}} \vert \mathrm{in}} \ee 
can be computed by Trotterizing the evolution operator and applying the quantum circuit shown in Fig.~\ref{fig:evo} $N \approx \frac{t}{\Delta t}$ times as well as operators $V_{\mathrm{in}}$ and $V_{\mathrm{out}}^\dagger$, before and after the evolution, respectively, where $\ket{\mathrm{in}} = V_{\mathrm{in}} \ket{\bm{0}}$ and $\ket{\mathrm{out}} = V_{\mathrm{out}} \ket{\bm{0}}$. Examples of quantum circuits implementing such operators are shown in Figures \ref{fig:QC0}, \ref{fig:QC1}, and \ref{fig:QC1b}.

\subsubsection{Two plaquettes}

\begin{figure}[ht!]
    \centering
\[\Qcircuit @C=3.2em @R=1em {   & \qw & \ctrl{4} & \qw & \qw & \qw  &\qw \\
    &\qw & \qw& \ctrl{4} & \qw &   \qw  &\qw  \\   &\qw& \qw  & \qw & \ctrl{4} & \qw   & \qw  \\  &\qw &\qw & \qw  & \qw &   \ctrl{4}   & \qw \inputgroupv{1}{4}{.8em}{1.8em}{\ket{\psi_{0}^{\text{QC}} (\alpha)}^{(1)}\ \ \ \ \ \ \ \ \ \ \ \ \ \ \ \ \ } \\    &\qw   & \ctrl{4}& \qw & \qw &   \qw    & \qw \\   &\qw  & \qw & \ctrl{3}  &\qw  & \qw  & \qw \\
  & \qw & \qw & \qw & \ctrl{2}   & \qw &\qw   \\ &\qw & \qw  & \qw & \qw  & \ctrl{1}  &\qw  \inputgroupv{5}{8}{.8em}{1.8em}{\ket{\psi_{0}^{\text{QC}} (\alpha)}^{(2)}\ \ \ \ \ \ \ \ \ \ \ \ \ \ \ \ \ } \\ \lstick{\ket{0}_{c}} &\gate{D(\xi_0)} & \gate{\widetilde{\mathcal{D}}(\xi)} & \gate{\widetilde{\mathcal{D}} (\xi)} & \gate{\widetilde{\mathcal{D}} (\xi)}   &\gate{\widetilde{\mathcal{D}} (\xi)}   &\measureD{n=1}   } \]
    \caption{Quantum circuit that generates the state Eq.~\eqref{eq:101}. It implements the trial state $\ket{\psi_{1}^{\text{QC}} (\alpha)}$ ( Eq.~\eqref{eq:97ab}) with $\xi_0 = \beta\xi$ and small $\xi$.}
    \label{fig:QC1b}
\end{figure}
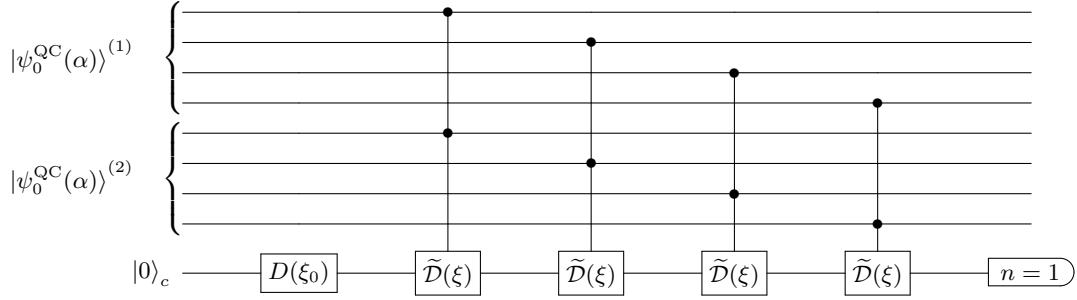

For two plaquettes with open boundary conditions, the Hamiltonian is given by Eq.~\eqref{eq:2pq1a}. It can be simulated with two qumode quadruplets labeled by $a=1,2$, of quadratures $\bm{q} (a) = (q_0(a),\vec{q} (a))$ and $\bm{p}(a) = (p_0(a),\vec{p}(a))$. The Hamiltonian can be written as
\begin{equation}
\label{eq:2pq1aQC}
    H^{\text{QC}} =  \frac{1}{2}\Big( [\vec{L} (1)]^2 +  [ \vec{K} (1) ]^2 + 2[\vec{L} (2) ]^2 + [ \vec{K} (2) ]^2    + \frac{1}{2} (\vec{K} (2) + 3\vec{L} (2)) \cdot (\vec{K} (1) - \vec{L} (1)) \Big) +\lambda(2-q_0(1) - q_0(2))
   \end{equation}
where $\vec{L}(a) = \vec{q}(a)\times \vec{p}(a)$ and $\vec{K}(a) = \vec{q}(a) p_0(a) - q_0(a) \vec{p}(a)$. To compute energy levels with a variational method, we adopt the coupled cluster ansatz
\be\label{eq:97ab} \braket{\bm{q}|\psi_{0}^{\text{QC}} (\alpha)} \propto \Phi(\alpha;\bm{q}(1))\Phi(\alpha; \bm{q}(2)) \ , \ \ \ket{\psi_{1}^{\text{QC}} (\alpha)} \propto (   \bm{q}(1)\cdot \bm{q} (2) - \beta) \ket{\psi_{0}^{\text{QC}} (\alpha)} \ee 
to be compared with the classical trial states Eq.~\eqref{eq:G2} and Eq.~\eqref{eq:G21}, for the ground and first-excited state, respectively. The parameter $\beta$ is fixed by demanding orthogonality of the two wavefunctions,
\be\label{eq:108b} \beta = \frac{\int d^4 q(1) d^4 q(2)\, \bm{q}(1)\cdot \bm{q} (2) [\Phi(\alpha;\bm{q}(1))\Phi(\alpha; \bm{q}(2))]^2}{\int d^4 q(1) d^4 q(2)\, [\Phi(\alpha;\bm{q}(1)) \Phi(\alpha; \bm{q}(2))]^2 } = \left(  \frac{I_2(2\alpha)}{I_1(2\alpha)} + \frac{\alpha}{4\Lambda^2} + \mathcal{O} (\frac{1}{\Lambda^4}) \right)^2\ee
A straightforward calculation confirms agreement with the classical expressions at leading order in $1/\Lambda$,
\be \epsilon_k^{\text{QC}}  = \bra{\psi_k^{\text{QC}} } H^{\text{QC}} \ket{\psi_k^{\text{QC}} } = \epsilon_k  + \mathcal{O} ( \frac{1}{\Lambda^2} ) \ , \ \ k=0,1 \ee
Engineering the trial state $\ket{\psi_{0}^{\text{QC}} (\alpha)}$ is straightforward, as it is the product of two copies of the single-plaquette trial state in Eq.~\eqref{eq:87}. Therefore, it can be implemented with two copies of the quantum circuit depicted in Figure \ref{fig:QC0}. For the trial state $\ket{\psi_{1}^{\text{QC}} (\alpha')}$, after implementing the circuit that creates $\ket{\psi_{0}^{\text{QC}} (\alpha')}$, we introduce an ancilla qumode of quadratures $(q^{(c)},p^{(c)})$ initialized in the coherent state $\ket{\xi_0}$ and entangle it with the 8 qumodes of $\ket{\psi_{0}^{\text{QC}} (\alpha')}$ by applying four controlled displacement gates with \textit{non-linear two-qumode control} targeting the ancilla qumode, collectively given by
\be\label{eq:110cc} \widetilde{\mathcal{D}} (\xi) = e^{i\xi \bm{q}(1)\cdot \bm{q} (2) p^{(c)} } \,.\ee
These are non-Gaussian gates that can be implemented with beam splitters and cubic phase gates. We have
\be e^{i\xi {q}_\mu (1) {q}_\mu (2) p^{(c)} } = U_{BS} e^{i\xi q_\mu^2 (1) p^{(c)}} U_{BS}^\dagger\,, \ee
where the gate $e^{i\xi q_\mu^2 (1) p^{(c)}}$ is implemented as in Eq.~\eqref{eq:79}. See Figure \ref{fig:QC2C} for the quantum circuit.

\begin{figure}
    \centering
    \includegraphics[width=0.42\linewidth]{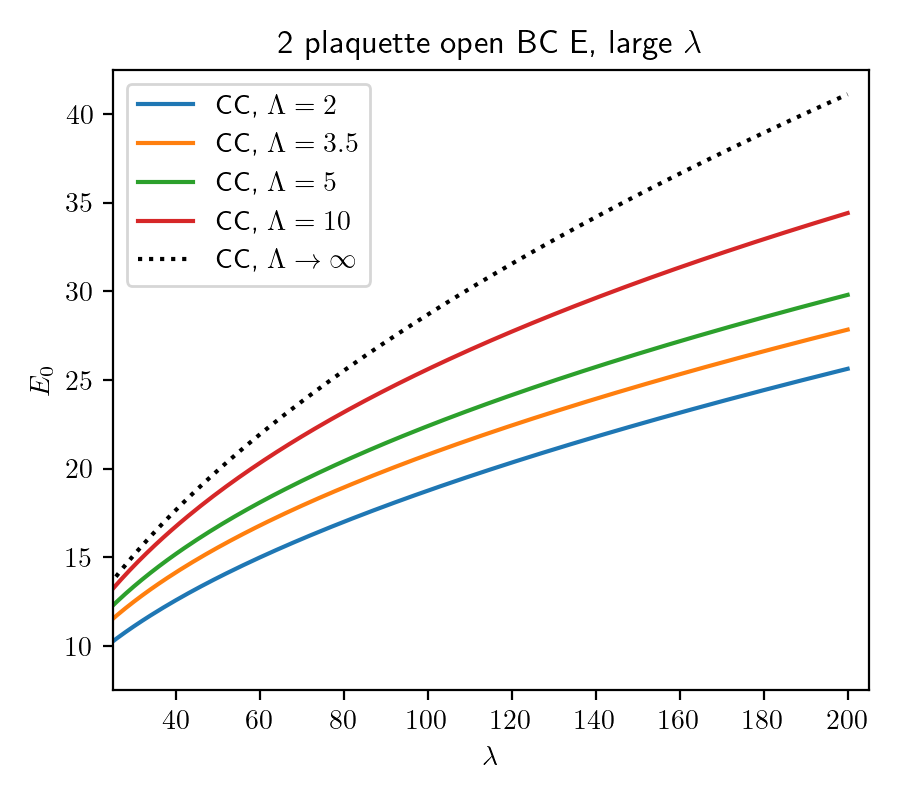}
    \includegraphics[width=0.42\linewidth]{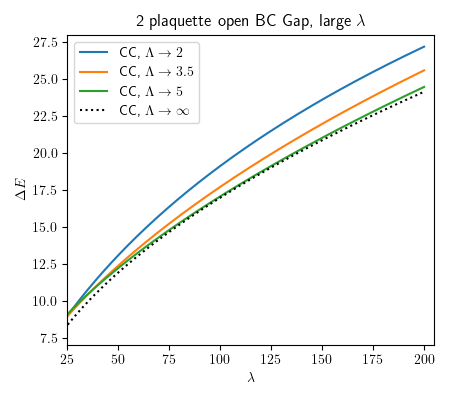}
    \caption{The ground state energy (left) and gap (right) of two plaquettes for large $\lambda$ using the coupled cluster (CC) ansatz \eqref{eq:97ab} for various values of the cutoff $\Lambda$.  }
    \label{fig:fig11}
\end{figure}

We then measure the ancilla qumode projecting it onto the single photon state. We arrive at the state
\be\label{eq:101} \left( \bm{q}(1)\cdot \bm{q} (2) - \frac{\xi_0}{\xi} \right) e^{-\frac{1}{2} [ \xi \bm{q}(1)\cdot \bm{q} (2) - \xi_0]^2} \ket{\psi_{0}^{\text{QC}} (\alpha')}\,.\ee
By choosing $\xi_0 = \beta \xi$ and a small $\xi$ (weak coherent  source for the ancilla qumode), this state is a good approximation to the desired state  $\ket{\psi_{1}^{\text{QC}} (\alpha')}$. The circuit that creates this state is shown in Figure \ref{fig:QC1b}, while the results of the quantum computation are shown in Figure \ref{fig:fig11}.

\begin{figure}[ht!]
    \centering
\[\Qcircuit @C=1.7em @R=1em {    \lstick{} & \push{(1)\mu\ \ } & \gate{F} & \multigate{1}{\text{BS}}   &\qw  &\qw  & \qw& \multigate{1}{\mathcal{K}'(-s)} &\qw& \qw & \link{1}{-1} &\qw & \multigate{1}{\text{BS}} &\qw&\link{1}{-1}&\gate{F^\dagger}&\qw\\
\lstick{} & \push{(1)\nu\ \ } &\qw &\ghost{\text{BS}} & \multigate{1}{\mathcal{K}'(s)} &\qw &\link{1}{-1} & \ghost{\mathcal{K}'(-s)} &\qw & \link{1}{-1} & \link{-1}{-1} & \multigate{1}{\mathcal{K}'(s)} & \ghost{\text{BS}}  &\qw&\link{-1}{-1}&\qw &\qw \\
\lstick{} & \push{(2)\rho\ \ } & \gate{F} & \multigate{1}{\text{BS}} & \ghost{\mathcal{K}'(s)} &\qw  &\link{-1}{-1}   & \multigate{1}{\mathcal{K}'(-s)} &\qw & \link{-1}{-1} & \link{1}{-1} & \ghost{\mathcal{K}'(s)} & \multigate{1}{\text{BS}} &\qw&\link{1}{-1}&\gate{F^\dagger}&\qw\\
\lstick{} & \push{(2)\sigma\ \ } &\qw &\ghost{\text{BS}} &\qw &\qw & \qw & \ghost{\mathcal{K}'(-s)} &\qw &\qw & \link{-1}{-1} & \qw & \ghost{\text{BS}}  &\qw&\link{-1}{-1}&\qw &\qw\inputgroupv{1}{4}{.75em}{3.25em}{\mathcal{U} (s) }\ \ \ \ \ \ \ \ \ \ \ \ \ \ }   \]
\caption{Quantum circuit for the evolution operator $\mathcal{U} (s) \equiv e^{-is J_{\mu\nu} (1) J_{\rho\sigma} (2)}$ contributing to the time evolution of the Hamiltonian Eq.~\eqref{eq:2pq1aQC} involving 4 qumodes.}
    \label{fig:evo2}
\end{figure}
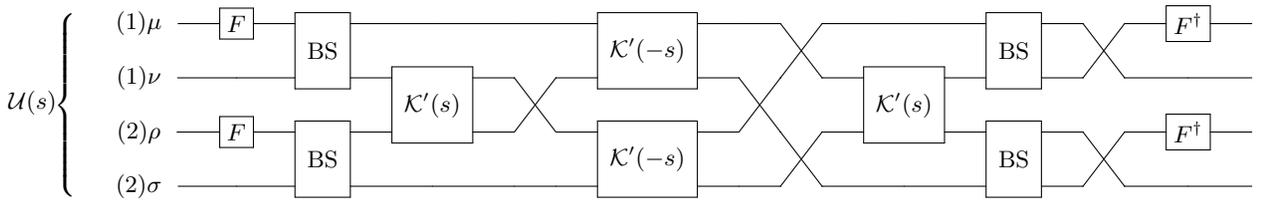

The time evolution is governed by the Hamiltonian Eq.~\eqref{eq:2pq1aQC}. It contains terms that involve a single qumode quadruplet, as in the single plaquette case discussed above, as well as terms that couple qumodes in different quadruplets, e.g., $ \vec{L} (1)\cdot \vec{L} (2)$. The latter can also be implemented with similar quantum circuits as the ones already introduced above. A typical term of the two-plaquette Hamiltonian in Eq.~\eqref{eq:2pq1aQC} involving qumodes from two quadruplets is $J_{\mu\nu}(1) J_{\rho\sigma}(2)$. Using Eq.~\eqref{eq:102J}, the time evolution with this part of the Hamiltonian can be implemented with non-Gaussian cross-Kerr gates using the quantum circuit depicted in Figure \ref{fig:evo2}.

\subsubsection{$N$ plaquettes}

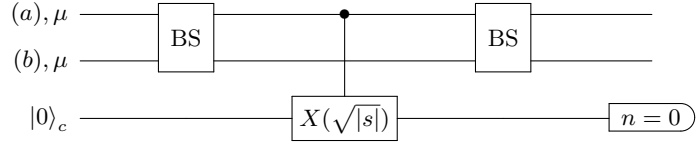
\begin{figure}[ht!]
    \centering
\[\Qcircuit @C=3.2em @R=1em { \lstick{(a), \mu}  & \multigate{1}{\text{BS}} & \ctrl{2} & \multigate{1}{\text{BS}} & \qw  \\
   \lstick{(b),\mu} &\ghost{\text{BS}} &\qw & \ghost{\text{BS}}  & \qw    \\ \lstick{\ket{0}_{c}} &\qw & \gate{X(\sqrt{|s|})}   &\qw &\measureD{n=0}   } \]
    \caption{Quantum circuit that generates a factor of the form $e^{-\frac{|s|}{2} ({q}_\mu (a) + {q}_\mu (b))^2}$ i.e., Eq.~\eqref{eq:103a}, which is required to construct the trial states in Eq.~\eqref{eq:97abNi}.}
    \label{fig:QC2}
\end{figure}

A system on a lattice with $N$ plaquettes can be simulated with $N$ qumode quadruplets of quadratures $(\bm{q} (a) ,\bm{p}(a))$ ($a=1,2,\dots, N$). $N$ grows linearly with the volume. The Hamiltonian is of the form
\be\label{eq:112H} H^{\text{QC}} = \frac{1}{2} \sum J_{\mu\nu} (a) J_{\rho\sigma} (b) + \lambda \left[ N - \sum_a q_0 (a) \right] \,,\ee
generalizing the 2-plaquette ($N=2$) Hamiltonian Eq.~\eqref{eq:2pq1aQC}. Energy levels can be estimated with a coupled cluster ansatz. Generalizing Eq.~\eqref{eq:97ab}, we may use the trial states
\be\label{eq:97abN} \braket{\bm{q}|\psi_{0}^{\text{QC}} (\alpha)} \propto \prod_a \Phi(\alpha;\bm{q}(a)) \ , \ \ \ket{\psi_{1}^{\text{QC}} (\alpha)} \propto \left( \sum_{a,b}  A_{ab} \bm{q}(a)\cdot \bm{q} (b) - \beta \right) \ket{\psi_{0}^{\text{QC}} (\alpha)} \,, \ee 
for the ground and first excited state, respectively. Here $\beta$ is determined by demanding orthogonality,
\be \beta = \sum_{a,b} A_{ab} \braket{{q}_0 (a)} \braket{{q}_0 (b)} \ , \ \ \braket{{q}_0} = \frac{\int d^4 q \ {q}_0 \ [\Phi(\alpha;\bm{q})]^2}{\int d^4 q  [\Phi(\alpha;\bm{q}) ]^2 } = \frac{I_2(2\alpha)}{I_1(2\alpha)} + \frac{\alpha}{4\Lambda^2} + \mathcal{O} (\frac{1}{\Lambda^4})\,, \ee
and the coefficients $A_{ab}$ are chosen to match the normal mode of the lowest frequency in the coupled harmonic oscillator approximation, which is valid in the large $\lambda$ limit (see  Eq.~\eqref{eq:57a}).

To engineer the trial ground state $\ket{\psi_0^{\text{QC}} (\alpha)}$, we need $N$ copies of the quantum circuit in Figure \ref{fig:QC0} which requires $\mathcal{O} (N)$ quantum resources. For the trial first excited state, we first create $\ket{\psi_0^{\text{QC}} (\alpha)}$, and then implement a string of unitaries of the form Eq.~\eqref{eq:110cc} (see the quantum circuit in Figure \ref{fig:QC1b}). We need $\mathcal{O} (N^2)$ quantum resources for the implementation of the quantum circuits in this case.

As pointed out in Section \ref{sec:V}, the ansatz in Eq.~\eqref{eq:97abN} fails to capture the vanishing behavior of the gap in the continuum limit. To remedy this, we introduce the modified trial states
\be\label{eq:97abNi} \ket{\varphi_{0}^{\text{QC}} (\alpha)} \propto e^{\alpha \sum_{ab} B_{ab} \bm{q} (a) \cdot \bm{q} (b)} \ket{\psi_{0}^{\text{QC}} (\alpha)} \ , \ \ \ket{\varphi_{1}^{\text{QC}} (\alpha)} \propto \left( \sum_{a,b}  A_{ab} \bm{q}(a)\cdot \bm{q} (b) - \beta \right) \ket{\varphi_{0}^{\text{QC}} (\alpha)}  \ee 
that simulate the ansatz Eq.~\eqref{eq:68psiN}. For this ansatz, we ought to generate additional factors of the form $e^{s\bm{q} (a) \cdot \bm{q} (b)}$. This can be accomplished by generating factors \be\label{eq:103a} e^{-\frac{|s|}{2} (\bm{q} (a) \pm \bm{q} (b))^2} = \prod_{\mu \in\{ 0,1,2,3\} } e^{-\frac{|s|}{2} ({q}_\mu (a) \pm {q}_\mu (b))^2} \ , \ee
where the sign is chosen opposite to the sign of the numerical coefficient $s$, since $\bm{q}^2(a)\approx 1$. Such factors can be engineered with the aid of an ancilla qumode, as discussed before, and beam splitters. The quantum circuit is shown in Figure \ref{fig:QC2}.

For the time evolution, after Trotterization, each term contributing to the Hamiltonian Eq.~\eqref{eq:112H} can be implemented independently for small time intervals. For the magnetic part of the Hamiltonian, we need to implement Gaussian displacement operators, $e^{i\lambda\Delta t q_0(a)}$, which results in a total of $N$ gates. All terms of the electric part of the Hamiltonian are of the form $e^{-is J_{\mu\nu} (a) J_{\rho\sigma} (b)}$ and can be implemented with quantum circuits of the form shown in Figure \ref{fig:evo2}. We need $\mathcal{O} (N^2)$ quantum resources to implement them. 

\subsection{QC with qumode triplets}

In this subsection, we discuss an alternative approach to QC with qumodes that avoids redundant degrees of freedom. Instead of using four independent qumodes to simulate an $SU(2)$ matrix, we can first solve the constraint $u_0^2 + \vec{u}^{\,2} =1$ to express $u_0$ in terms of the other three components and then use a triplet of qumodes to simulate $\vec{u}$. Apart from utilizing fewer resources, this approach avoids having to introduce a factor of the form Eq.~\eqref{eq:85} restricting the support of the wavefunction. An additional important advantage is that it simulates the system at large-$\lambda$ efficiently because, in that regime, the system reduces to a set of coupled three-dimensional harmonic oscillators.  The energy gap in the large $\lambda$ limit can be computed exactly without having to resort to a variational method.

The drawback is that as one goes away from large $\lambda$ limit, the Hamiltonian is no longer polynomial in the quadratures. To address this, one has to expand the non-polynomial expressions, e.g., $u_0 = \sqrt{1 - \vec{u}^{\,2}} = 1 - \frac{1}{2} \vec{u}^{\,2} + \dots$, and truncate the expansion. This is equivalent to a weak-coupling expansion (in $1/\sqrt{\lambda} \sim g^2$). As we have discussed, at leading order the system reduces to a set of coupled three-dimensional harmonic oscillators. It is conveniently described by its normal modes. When higher-order terms are included in the Hamiltonian, it is easier to express them in terms of the original coordinates of the coupled harmonic oscillators. It is therefore important to implement the transformation $\mathcal{G}$ between coupled harmonic oscillators and normal modes efficiently.

For a single plaquette, there is only one harmonic oscillator, so trivially, $\mathcal{G} = \mathbb{I}$. For two plaquettes, the transformation is implemented with a 50:50 beam splitter, $\mathcal{G} = U_{BS}$. In general, for $N$ plaquettes, $\mathcal{G}$ can be implemented efficiently with $N \times N$ interferometers and single-mode squeezers, according to the Bloch-Messiah singular-value decomposition~\cite{Braunstein2005,Marshall:2015mna,Briceno:2023xcm}.

The computation of matrix elements of the time evolution operator $e^{-itH}$ are less problematic with triplets because no unphysical parameter, like $\Lambda$, needs to be introduced and, hence, the error is under better control. Another advantage of this approach is that it better captures the next-to-leading-order contribution to the gap. This is important because at leading order the gap vanishes in the large volume limit, whereas next-to-leading-order corrections yield a finite gap~\cite{Muller:1983rs, Feynman1981}. We will discuss this approach further in future work.

\section{Summary and future directions}\label{sec:VII} 

In this work, we introduced a novel approach to study $SU(2)$ lattice gauge theory using continuous-variable (CV) quantum computation. We validated this quantum computational framework by comparing it against known theoretical results.
We started by implementing the CV quantum computation to solve the dynamics of a single $SU(2)$ plaquette. Using qumodes to represent the gauge degrees of freedom, we calculated the energy spectrum and validated it against exact analytical results. This demonstrates that the CV quantum approach accurately captures the gauge theory's essential dynamics and energy structure. We extended the method to a ladder of $N$ plaquettes and a two-dimensional grid of $N\times N$ plaquettes, for which we computed the ground states and energy gaps. We benchmarked these results against traditional methods, such as the weak coupling approximation and exact diagonalization. The quantum simulations showed good agreement with known results, validating the use of CV qumodes to simulate multi-plaquette systems. We discussed two distinct quantum computational approaches for encoding the $SU(2)$ gauge group into qumodes: Using either quadruplets or triplets of qumodes per physical link. These quantum computational methods may offer a scalable and efficient path for simulating more complex gauge theories. 
 
Our work provides a foundation for using CV quantum computation to study SU(2) lattice gauge theory, offering a promising direction for tackling the complexities of gauge theories in $3+1$ dimensions. The ability to encode the infinite degrees of freedom of gauge fields into a CV framework reduces the challenge of digitizing the theory, particularly for problems that require capturing real-time dynamics or strong coupling regimes where traditional Monte Carlo methods struggle. As quantum hardware advances toward fault tolerance, CV quantum computing may become an important tool for studying non-Abelian gauge theories, enabling precise simulation of phenomena like quark confinement, topological phases, and high-energy scattering processes in QCD.

The extension of our approach to more complex gauge groups, such as $SU(3)$, holds significant potential. This would open the door to quantum simulation of QCD and could provide new insights into the behavior of the strong interaction. An alternative interesting route to $SU(3)$ is to consider irreducible Schwinger bosons, as discussed in Ref.~\cite{Anishetty:2009ai}. Additionally, incorporating tensor network techniques and hybrid classical-quantum algorithms could further optimize the method for near-term quantum devices. As quantum computing technology continues to evolve, the continuous-variable framework may play a critical role in addressing open problems in particle physics, such as exploring the real-time dynamics of gauge theories, thermalization processes, and out-of-equilibrium phenomena in subatomic physics.

Future work will also focus on refining error mitigation techniques specific to CV systems and the exploration of hybrid approaches that combine discrete variables (qubits) and CV substrates (qumodes)~\cite{Davoudi:2021ney,Crane:2024tlj,Araz:2024kkg}. The ultimate goal will be to scale these methods using fault-tolerant quantum computers allowing for the simulation of real-world quantum field theories.

\acknowledgments
We thank Shane Thompson for useful discussions and collaboration during the early stages of this work.  We acknowledge support by DOE ASCR funding under the Quantum Computing Application Teams Program, NSF award DGE-2152168, and DOE Office of Nuclear Physics, Quantum Horizons Program, award DE-SC0023687. This research used resources of the Oak Ridge Leadership Computing Facility, which is a DOE Office of Science User Facility supported under Contract DE-AC05-00OR22725. This material is based upon work supported by the U.S. Department of Energy, Office of Science, Contract No. DE-AC05-06OR23177, under which Jefferson Science Associates, LLC operates Jefferson Lab. The research was also supported by the U.S. Department of Energy, Office of Science, National Quantum Information Science Research Centers, Co-design Center for Quantum Advantage under contract number DE-SC0012704 and the DOE, Office of Science, Office of Nuclear Physics, Early Career Program under contract No~DE-SC0024358.


\providecommand{\href}[2]{#2}\begingroup\raggedright\endgroup

\end{document}